\documentclass{aa}  
\usepackage{graphicx}
\usepackage{siunitx}
\DeclareSIUnit \parsec {pc}
\DeclareSIUnit\year {yr}
\usepackage{txfonts}
\usepackage{multirow}
\usepackage{booktabs}
\usepackage[utf8]{inputenc}
\usepackage{newunicodechar} 
\usepackage{amsmath}
\usepackage{bm}
\newunicodechar{⌀}{\ensuremath{\diameter}}
\DeclareUnicodeCharacter{00A0}{~}	
\usepackage{natbib}
\bibpunct{(}{)}{;}{a}{}{,} 
\usepackage[colorlinks=true, linkcolor = blue, urlcolor=blue, citecolor = blue]{hyperref}
\usepackage{txfonts}

\newcommand*{\rom}[1]{\expandafter\@slowromancap\romannumeral #1@}
\begin{document}

   \title{Tracing the evolutionary pathways of dust and cold gas in high-z quiescent galaxies with \texttt{SIMBA}}

   \author{G. Lorenzon
          \inst{1}
          \and
          D. Donevski\inst{1,2}
          \and K. Lisiecki \inst{1}	
       \and C. Lovell\inst{3}
       \and M. Romano\inst{1,4}
	\and D. Narayanan\inst{5,6}
	\and R. Dav\'e\inst{7,8}
    \and A. Man\inst{9}
    \and K. E. Whitaker\inst{10}
    \and A.~Nanni\inst{1,11}
    \and A. Long \inst{12,13}
    \and M. M. Lee \inst{14}
    \and Junais \inst{1}
    \and K. Małek \inst{1}
    \and G. Rodighiero\inst{15,4}
    \and Q. Li\inst{16}
   }

   \institute{National Center for Nuclear Research, Pasteura 7, 02-093 Warsaw, Poland\\
              \email{giuliano.lorenzon@ncbj.gov.pl}
         \and 
    SISSA, Via Bonomea 265, 34136 Trieste, Italy
          \and
    Institute of Cosmology and Gravitation, University of Portsmouth, Burnaby Road, Portsmouth PO1 3FX, UK
          \and
    INAF, OAPD, Vicolo dell'Osservatorio, 5, 35122 Padova, Italy
	\and
    Department of Astronomy, University of Florida, 211 Bryant Space Sciences Center, Gainesville, FL, USA
	\and
	Cosmic Dawn Center (DAWN), Copenhagen, Denmark
	\and 
	Institute for Astronomy, Royal Observatory, University of Edinburgh, Edinburgh EH9 3HJ, UK
	\and
    Department of Physics \& Astronomy, University of the Western Cape, Robert Sobukwe Rd, Bellville, 7535, South Africa
	\and
	Department of Physics \& Astronomy, University of British Columbia, 6224 Agricultural Road, Vancouver, BC V6T 1Z1, Canada
        \and
    Department of Astronomy, University of Massachusetts, Amherst, MA 01003, USA
        \and
    INAF - Osservatorio astronomico d'Abruzzo, Via Maggini SNC, 64100, Teramo, Italy
        \and 
    Department of Astronomy, University of Washington, Seattle, WA 98195-1700, USA
    \and
    Department of Astronomy, The University of Texas at Austin, 2515 Speedway Boulevard Stop C1400, Austin, TX 78712, USA
	\and
    DTU-Space, Technical University of Denmark, Elektrovej 327, DK2800 Kgs. Lyngby, Denmark
        \and
    Dipartimento di Fisica e Astronomia, Università di Padova, Vicolo dell'Osservatorio, 3, I-35122 Padova, Italy
        \and
    Max-Planck-Institute für Astrophysik, Karl-Schwarzschild-Strasse 1, 85740, Garching, Germany}
   \date{}


  \abstract
{Recent discoveries of copious amounts of dust in quiescent galaxies (QGs) at high redshifts ($z\gtrsim 1-2$) challenge the conventional view that these objects have negligible interstellar medium (ISM) in proportion to their stellar mass. We make use of the \texttt{SIMBA} hydrodynamic cosmological simulation to explore how dust and cold gas evolve in QGs in relation to the quenching processes affecting them. 
We apply a novel method for tracking the changes in the ISM dust abundance across the evolutionary history of QGs identified at $0 < z \lesssim2$ in both cluster and field environments.
The QGs transition from a diversity of quenching pathways, both rapid and slow, and exhibit a wide range of times elapsed between the quenching event and cold gas removal (from $\sim650$ Myr to $\sim8$ Gyr). Contrary to some claims, we find that quenching modes attributed to the feedback from active galactic nuclei (AGN) do not affect dust and cold gas within the same timescales. 
Remarkably, QGs may replenish their dust content in the quenched phase primarily due to internal processes and marginally by external factors such as minor mergers. Prolonged grain growth on gas-phase metals appears as the key mechanism for dust re-formation, being effective within $\sim100$ Myr after the quenching event, and rapidly increasing the dust-to-gas mass ratio in QGs above the standard values ($\delta_{\rm DGR}\gtrsim1/100$). Consequently, despite heavily depleted cold gas reservoirs, roughly half of QGs maintain little evolution of their ISM dust with stellar age within the first 2 Gyr following the quenching. Overall, we predict that relatively dusty QGs ($M_{\rm dust}/M_{\star}\gtrsim10^{-3}-10^{-4}$) arise from both fast and slow quenchers, and are prevalent in quenched systems of intermediate and low stellar masses ($9<\log(M_{\star}/M_{\odot})<10.5$). This strong prediction poses an immediate quest for observational synergy between e.g., James Webb Space Telescope (JWST) and the Atacama Large Millimeter Array (ALMA).}   


\maketitle
%

\section{Introduction}
\label{Introduction}
The passively evolved, so-called quiescent galaxies (QGs), are often characterised by old stellar populations and a very low or almost absent level of ongoing star-formation activity. These features place QGs below the star-forming main-sequence (MS; e.g., \citealt{daddi05}, \citealt{toft07}, \citealt{dokkum10}, \citealt{rodighiero2011lesser}, \citealt{schreiber15}). 
There are different paths that star-forming galaxies (SFGs) can take to reach the quiescent stage as QGs both from a theoretical (e.g., \citealt{feldmann15a}, \citealt{rodriguez2019mergers}, \citealt{zheng2022rapidly}) and observational (e.g., \citealt{faber07}, \citealt{kriek2008detection}, \citealt{peng10}, \citealt{Tacchella_2022}, \citealt{akhshik23}) perspective. The conditions required to undergo the different evolutionary processes, which seem to be mainly connected with the timescale of the quenching phase, are still poorly understood  (e.g., \citealt{park2023rapid}, \citealt{corcho2023ageing}).
The quenching time and the global properties of the QGs are affected by internal mechanisms, such as active galactic nuclei (AGN; e.g., \citealt{crenshaw03}, \citealt{ma2022effects}, \citealt{stacey22}), star formation activity, and external causes such as ram pressure stripping (e.g., \citealt{Gott1972}, \citealt{fossati2020}, \cite{junais2022virgo}), starvation (e.g., \citealt{vandevoort2016eaglestarvation}, \citealt{maier19}), and harassment (e.g., \citealt{merritt1983harassment}, \citealt{Moore_1998}). 

Whatever physical mechanism is in place, one of the important consequences of quenching the star formation is the evolution of its agents – gas and dust in the cold interstellar medium (ISM). While cold $H_2$ gas is the fundamental ingredient for the star formation, dust is critical in the thermal balance of the gas and in the formation of $H_2$ molecules (e.g., \citealt{cuppen2017grain}). Metals in the gas phase are thought to support the growth of dust particles (\citealt{zhukovska2016modeling}), in turn regulating both the star formation and its quenching. Dust further affects the spectral energy distributions (SED) of galaxies, by absorbing the stellar light at shorter wavelengths, and re-emitting it in the far-infrared (FIR). After the star formation ceases, both environmental and internal processes continue to affect ISM evolution, but the quiescent phase of a galaxy has often been associated with a period of simple and predictable evolution. However, recent observations suggest that several processes can still happen within the remaining cold ISM. Namely, various studies  unveil anomalously large dust and/or gas content (\citealt{gobat2018unexpectedly}, \citealt{spilker2018molecular}, \citealt{magdis2021interstellar}, \citealt{belli21}, \citealt{kalita2021ancient}, \citealt{morishita22}, \citealt{donevski23}, \citealt{lee2023high}, \citealt{kakimoto24}, \citealt{setton24}), incompatible with a purely passive evolution. Consequently, quantifying changes in the ISM dust after the cessation of star formation has become one of the key open questions of galaxy formation and evolution.

Despite its importance, very little is known about the evolution of dust and cold gas in QGs, in stark contrast to SFGs, for which a reasonably good census of the dust and cold gas content has been established up to very high redshifts ($z\sim6$; \citealt{liu19b}, \citealt{donevski20}, \citealt{hodge2020high}, \citealt{gruppioni2020alpine}, \citealt{kokorev21}, \citealt{inami2022alma}). The main challenge in studying the ISM component of passive galaxies is the faint FIR emission of QGs, and the difficulty in spectroscopically inferring their cold gas masses ($M_{\rm H_{2}}$) and gas phase metallicities ($Z_{\rm gas}$), mainly due to the absence of strong emission lines. However, recent observational results have shown great promise, either by stacking the dust continuum of QGs (\citealt{man16}, \citealt{gobat20}, \citealt{millard2020s2cosmos}, \citealt{magdis2021interstellar}, \citealt{blanquez2023gas}), or by large statistical studies of individually detected QGs in deep fields (\citealt{donevski23}). Moreover, a handful of recent works utilised extensive ALMA sub-millimeter follow up of individual QGs in order to attempt to measure $M_{H_2}$ of QGs in the distant Universe ($z>1-3$). This was done either through dust continuum (\citealt{whitaker21b}, \citealt{gobat2022uncertain}, \citealt{sase23a}, \citealt{lee2023high}), CO lines (\citealt{williams21}, \citealt{belli21}, \citealt{morishita22}) or atomic carbon emission at 158 $\mu$m ($\left[\rm CII\right]$, \citealt{deugenio23}).

These studies reported that a non-negligible fraction of massive ($10< \log(M_{\star}/M_{\odot})<11$) QGs at $0.1<z<2.5$ may be fairly dusty, with average specific dust masses $f_{\rm dust} = M_{\rm dust}/M_{\star} \gtrsim 10^{-4}$, systematically lower than in dusty star-forming galaxies (DSFGs) at these redshifts. However, these studies also reported few conflicting conclusions regarding dust evolution in QGs. (1) While stacking studies anticipate a strong anti-correlation between dust content and stellar age, statistical studies based on individual detections found a shallower trend accompanied by a large scatter ($>$2 orders of magnitude) in $f_{\rm dust}$ at a given stellar age, $M_{\star}$ and/or $Z_{\rm gas}$ (\citealt{li19}, \citealt{morishita22}, \citealt{donevski23}, \citealt{lee2023high}). The latter finding points towards complex and evolving ISM conditions, and diversity in the quenching mechanisms and/or dust formation pathways. (2) Due to the difficulty in constraining both $M_{\rm dust}$ and $M_{\rm H_{2}}$, it remains unclear whether QGs obey the canonical dust-to-gas mass ratio ($\delta_{\rm DGR}\sim 1/100$) seen in local SFGs, or if they cover a much wider range of values (e.g., \citealt{whitaker21a}, \citealt{michalowski23}). (3) There is no consensus on whether the dust content observed in QGs is of internal origin (owing to past star-formation), or if external processes (i.e., mergers, accretion of satellites), have an influential role. (4) Lastly, it is debatable how long the dust and cold $\rm H_{2}$ gas will persist after the quenching. Recent studies suggest either a quick depletion of dust and gas ($\lesssim 100-500$ Myr;  \citealt{sargent2015direct}, \citealt{williams21}, \citealt{whitaker21b}), whereas others report a possibility that timescales for ISM removal may be relatively long ($\gtrsim 1$ Gyr; \citealt{gobat2018unexpectedly}, \citealt{michalowski2019fate}, \citealt{woodrum2022molecular}).

The advent of the James Webb Space Telescope (JWST) has led to the identification of QGs up to very high redshifts ($z\sim3-7$, \citealt{carnall2023surprising}, \citealt{looser23}), including few candidate dusty QGs (e.g., \citealt{rodighieroletter2022}, \citealt{setton2024uncover}), suggesting new and interesting scenarios for the rapid evolution of galaxies in the young Universe. It is then essential to identify and study QGs soon after the quenching episode, as we can then test directly the relative contribution of different quenching processes on the evolution of the cold dust and gas. Because of the many observational challenges outlined above, cosmological simulations of galaxy evolution provide a unique tool for addressing these questions (e.g., \citealt{Wright2019}, \citealt{appleby2020impact}, \citealt{donnari21}). In this work we utilise the state-of-the-art cosmological simulation \texttt{SIMBA}, which tracks the dust life cycle in a self consistent way \citep{simba2019}, meaning that dust grains form, grow and get destroyed accounting for the conditions of the ISM within the simulation itself, evolving together with the galaxy. Our goal is to provide the framework for quantifying changes of the dust and cold gas in QGs during the different stages of their evolution. To this aim, the present study plans to give viable answers to key open questions, namely (1) What physical mechanisms help replenish the dust reservoirs in QGs after the cessation of star formation?; (2) What is the link between the co-evolution of the dust and cold gas in QGs and the quenching processes affecting them?; (3) What is the role of the environment in shaping the fate of ISM in QGs?

We organise this work as follows: in Sect.~\ref{The SIMBA simulation} we present the \texttt{SIMBA} simulation, focusing on the prescriptions for dust evolution and stellar and AGN feedback. Section~\ref{Method} presents the criteria for the selection of passive galaxies and their environment, defines different representative points of the galaxy evolution and describes the main parameters explored in the analysis. Section~\ref{Results} describes the predicted global evolution of dust-related properties in QGs as well as their dependence with the environment. In Sect.~\ref{discussion} we provide extensive discussion on the scenarios responsible for the observed diverse evolutionary pathways in QGs.
A summary of this work, and possible implications for observational studies, are presented in Sect.~\ref{conclusion}.

Throughout this work adopt the same $\Lambda$-CDM cosmology of the \texttt{SIMBA} simulation, that is $\Omega_{\rm m} = 0.3$, $\Omega_{\Lambda} = 0.7$, $\Omega_{\rm b} = 0.048$, $H_0 = $ \SI{68}{\kilo\meter\per\second\per\mega\parsec}. We adopt the same cosmology used in \citealp{speagle2014highly} when calculating the star forming main sequence (MS) therein, for comparison with our data sample. We assume a Chabrier initial mass function (IMF; \citealt{chabrier03}) consistent with \texttt{SIMBA}.

\section{The \texttt{SIMBA} simulation} \label{The SIMBA simulation}

\texttt{SIMBA} is a state-of-the-art cosmological hydrodynamic simulation that evolves together dark and baryonic matter in a self-consistent way. \texttt{SIMBA} is perfectly suited for our science goal due to its careful treatment of stellar and AGN feedback on sub-galactic scales and active dust modelling. \texttt{SIMBA} allows for a self-consistent evaluation of both dust destruction and dust growth rates at each evolutionary step, together with a rich chemical evolution of the ISM, and a more realistic treatment of AGN feedback with respect to previous iterations, such as \texttt{MUFASA} \citep{mufasa2016}. 

We make use of the full-physics (m100n1024) simulation that consists of a box of side $100\:\rm h^{-1}Mpc$ containing $2\times 1024^{3}$ particles (meaning $1024^{3}$ dark matter particles and $1024^{3}$ baryon particles). The minimum stellar mass of a resolved galaxy is $5.8\times10^8\:\rm M_{\odot}$, while the minimum initial mass for a gas particle is $1.82\times10^7\:\rm M_{\odot}$ and it is $9.6\times10^7\:\rm M_{\odot}$ for a dark matter particle. The minimum gravitational softening length is $0.5\:\rm h^{-1}Mpc$. The entire run covers the redshift range $0 < z < 249$. Below we describe the major recipes used to model the feedback and dust physics in \texttt{SIMBA}. For a more exhaustive description of \texttt{SIMBA} we refer the reader to \cite{simba2019}. 

\subsection{Star formation}

Star formation in \texttt{SIMBA} is based on the molecular gas fraction following \cite{schmidt1959rate} law with a fixed star formation efficiency $\epsilon_{\star} = 0.02$ from \cite{kennicutt1998global}. The SFR is proportional to the gas density and molecular gas fraction, and inversely scale with the local dynamical time. The molecular gas fraction at a sub-grid level is estimated using the \cite{krumholz2009atomic} in order to compensate for the limitations of the simulation's resolution. The estimation assumes a simple model for the photodissociation of molecular gas in finite clouds (\cite{krumholz2009star}). In the model, the ability of a cloud of molecular gas to self-shield against the interstellar radiation field is strongly dependent on the gas column density and weakly on the metallicity. 

\subsection{Feedback physics}

There are two main categories of feedback in \texttt{SIMBA}: stellar-driven and AGN-driven. For the stellar driven mechanism we distinguish between instantaneous feedback from young massive stars and Type II supernovae (SNe), and a delayed feedback from evolved stars and Type $\rm Ia$ SNe. Black holes (BHs) are seeded and grown during the simulation using torque-limited (momentum-driven infall in the BH through an accretion disk) and Bondi (isotropic infall) accretion modes. The accretion energy drives feedback that acts to quench galaxies. The AGN driven mechanism, coming from the black hole accretion, is tuned to mimic the observed dichotomy in black hole growth mode (\citealp{Heckman2014}) and the observed mode of energy injection into the large scale gas component. In \texttt{SIMBA} this is obtained by dividing the kinetic feedback into a radiative-mode and a jet-mode, characterized by different regimes of the Eddington ratio $f_{\rm Edd} = \dot{M}_{\rm BH}/\dot{M}_{\rm Edd}$, where $\dot{M}_{\rm BH}$ is the total BH accretion rate (Bondi plus Torque-limited) and $\dot{M}_{\rm Edd}$ is the Eddington accretion rate (accretion rate for which the BH emit at Eddington luminosity). Values of $f_{\rm Edd} > 0.2$ correspond to a radiative mode, where mass-loaded winds are ejected radially with velocity of the order of $10^3\:\rm km\:s^{-1}$. For $f_{\rm Edd} < 0.02$ the feedback is a fully jet mode, with hot gas expelled through strictly collimated bipolar jets with zero initial aperture and velocities capped at $\SI{7e3}{\kilo\meter\per\second}$. The transition regime is characterized by winds of increasing velocity as $f_{\rm Edd}$ decreases to jet mode. A lower mass limit for the triggering of the jet mode is set to $M_{\rm BH} = 10^{7.5} M_{\odot}$ based on considerations from radio observations \citep{barisic2017}. An additional X-ray feedback heats the gas surrounding the AGN following \cite{choi2012} when the jet mode is active. The heating is applied directly to the non-ISM gas, deposited as both heat and a radial outwards kick for the ISM gas.

\subsection{ISM dust in \texttt{SIMBA}} \label{Interstellar medium}

In \texttt{SIMBA} the ISM evolves naturally from the initial conditions, accounting for the  production, destruction, and transport of its various components through feedback mechanisms, thus ensuring self-consistence. The complete description of the dust framework that models the production, growth and destruction of dust grains in \texttt{SIMBA} is introduced in \cite{liqi19}. The baryonic component of a galaxy is initially in gas phase, and then it is turned into stellar component, providing energy injection by modeling the net effect from the unresolved contribution of SNe shocks, radiation pressure, and stellar winds. A spatially uniform photoionizing background is used in the context of gas heating/cooling. The action of both SNe and winds from asymptotic giant branch (AGB) stars are used to update the metallicity value of nearby gas particles. This tracks 9 elements, with good approximation of the whole metal budget: C, N, O, Ne, Mg, Si, S, Ca, Fe. The yields adopted for Type II SNe are from \cite{iwamoto1999nucleosynthesis}, while for Type $\rm Ia$ SNe and AGB stars, those from \cite{oppenheimer2008mass} are used. 

The dust model within \texttt{SIMBA} assumes that dust is produced by condensation of a fraction of metals from Type II SNe and AGB ejecta. The
condensation efficiencies for AGB stars and and Type II SNe are
based on the theoretical models of \cite{ferrarotti06} and \citep{bianchi07}, respectively. The model does not include a contribution from Type Ia SNe, as opposed to some other studies that adopt the same condensation efficiency between Type $\rm Ia$ SNe and Type II SNe (e.g., \citealt{popping17}). Metals can be locked into dust grains on-the-fly or expelled from the ISM by means of metal-loaded stellar winds. A self-consistent treatment of dust allows grains to grow and be destroyed depending on the energy injection from the unresolved SNe activity and AGN activity. For simplicity, dust grains are assumed to have the same initial size at formation, that is of radius $r = $ \SI{0.1}{\micro\meter} and to follow directly the gas dynamics. 

Dust growth is based on a modified version of the \cite{dwek1998evolution} prescription, deriving the mass of produced dust from the mass of different metals and accounting for the chemical composition and type of the stellar population seeding the ISM. Dust mass can then increase due to accretion in the ISM, based on a two-body collision of grains. This is computed on-the-fly, based on mass, density and temperature of metal-enriched gas \citep{McKinnon2017}.  

Destruction of grains can happen by thermal sputtering \citep{1995Tsai}, violent action of SNe shocks \citep{Hopkins2015} or due to astration (consumption by star formation). In the first case the rate of destruction depends on the temperature, while in the second case it depends on the mass of the shocked gas and on a parameter $\epsilon$ representing the destruction efficiency, since shocks are not resolved in \texttt{SIMBA}. Moreover, dust grains can be destroyed instantaneously if they reside in regions of gas impacted by the X-ray-mode AGN feedback, jets or hot winds, while outflow mechanisms such as the radiative-mode AGN feedback and the stellar feedback can heat-up and move dust out of the galaxy.

We restrict our results to the redshift range $0\lesssim z\lesssim2$ in order to compare it with observationally derived statistics around the cosmic noon. It is worth noting that the \texttt{SIMBA} flagship run results in $M_{\rm dust}$ functions in a good agreement with observations at $z < 2$ (\citealt{liqi19}). It is also consistent with the observed sub-millimeter number counts up to $z\sim2$ (\citealt{donevski20}, \citealt{lovell21}) and even with the observed dust fractions in QGs at $z<1$ (\citealt{donevski23}).

\section{Methods} \label{Method}

\subsection{Identification of quenched galaxies} \label{Identification of quenched galaxies}
   
For identifying QGs in the range $0\lesssim z \lesssim 2$ we adopt a redshift-dependent parametrisation from \cite{Pacifici_2016} which defines both a threshold for the star-forming phase and a threshold for the quenched phase. Such thresholds are based on the instantaneous specific SFR (sSFR; $\rm sSFR \equiv \rm SFR/M_{\star}$), with SFGs having $\rm sSFR > 1/\tau (z)$ and QGs having $\rm sSFR < 0.2/\tau (z)$, with $\tau (z)$ being the age of the Universe at redshift $z$. The thresholds have empirical origin, coming from the observed evolution of the normalisation of the MS (i.e., \citealt{2012Whitaker}, \citealt{2014Fumagalli}, \citealt{2012Whitaker})\footnote{The SF threshold roughly corresponds to the evolution of the level of the sSFR in the MS minus 0.3 dex, which is generally considered to be the lower limit of the MS scatter.
The quenching threshold roughly corresponds to the average observed distance of at least 0.4 dex below the MS, ensuring that the galaxy can be considered quenched independently of its mass.}. We ensure that this method provides a great agreement with a selection based on the distance to the MS \citep{speagle2014highly}, as validated in previous \texttt{SIMBA} works on QGs (\citealt{appleby2020impact}, \citealt{akins2022quenching}).

In Fig.~\ref{fig:SFH_examples} we examplify six representative star formation histories (SFH) of \texttt{SIMBA} galaxies that quench at $z\lesssim 2$. The region between the intersection of the SFH with the two thresholds is highlighted in green. Fig.~\ref{fig:SFH_examples} outlines variety of SFHs for the displayed QGs. It is noticeable that SFHs are not always smooth, and can exhibit large oscillations between subsequent evolutionary points. Due to this irregular behavior, the sSFR of the galaxy can temporary fall below the quenching threshold and immediately rise again above it, remaining in a state of low-but-consistent star formation for a long time. Such events must not be accounted for as quenching events since they are ambiguous in nature. We then follow the prescription of \cite{rodriguez2019mergers} to avoid such cases by interpolating the SFH with a cubic B-Spline \citep{dierckx1975algorithm}. The point at which the SFH drops below the SF and the quenchign thresholds is calculate as the intersection of the functions $\rm sSFR \equiv 1/\tau (z)$ and $\rm sSFR \equiv 0.2/\tau (z)$ respectively with the interpolated SFH. We further request that the sSFR remains below the star forming threshold for a minimum time of $0.2 \times \tau (z_q)$,  where $\tau (z_q)$ is the cosmic time at the redshift $\rm z_q$ of the candidate quenching event. 

A galaxy can also have multiple quenching events if it is rejuvenated after crossing the quenching threshold. Rejuvenation can happen for different reasons, usually involving the formation of a cold gas region through inflows of molecular hydrogen (\citealp{young2014}) because of morphological changes, environmental effects or merging events (\citealp{rowlands2018galaxy}, \citealp{woodrum2022molecular}, \citealp{chauke2018star}, \citealp{zhang2023simple}). It is then important to evaluate these different quenching phases in order to connect them to their respective possible causes. However, as noted in Sect.~\ref{Mergers} the fraction of rejuvenated galaxies found in our sample appears to be fairly low, so we can neglect the effects of multiple quenching in our statistical treatment. We select in the specific case of multiple quenching only the first quenching event. The quenching time is calculated as the cosmic time interval that elapsed between the quenching event (at $z_{\rm q}$) and the crossing of the star forming phase threshold previous to that (at $z_{\rm sf}$): 

\begin{equation}
  t_{\rm quench} = \tau (z_{\rm q}) - \tau (z_{\rm sf}).
  \label{eq:t_quench}
\end{equation}

As we study the redshift range $0 < z \lesssim 2$, we exclude from our analysis all galaxies that do not cross quenching threshold in this range. In our accompanying work (Lorenzon et al., in prep.) we present the analysis of the effect of formation redshifts to global ISM properties for all \texttt{SIMBA} galaxies.

\subsection{Evolutionary phases and timescales} \label{Evolution stages}

The use of a cosmological simulation allows a simultaneous study from both a statistical and single-case perspective. We link these two aspects by comparing different galaxies during a few significant phases of their evolution. In this way, we can determine the different SFH that galaxies experience to reach the same stage in their evolution, and monitor which properties are influencing such paths. 

\begin{figure}[tbh!]
\centering
\includegraphics[width=0.92\columnwidth]{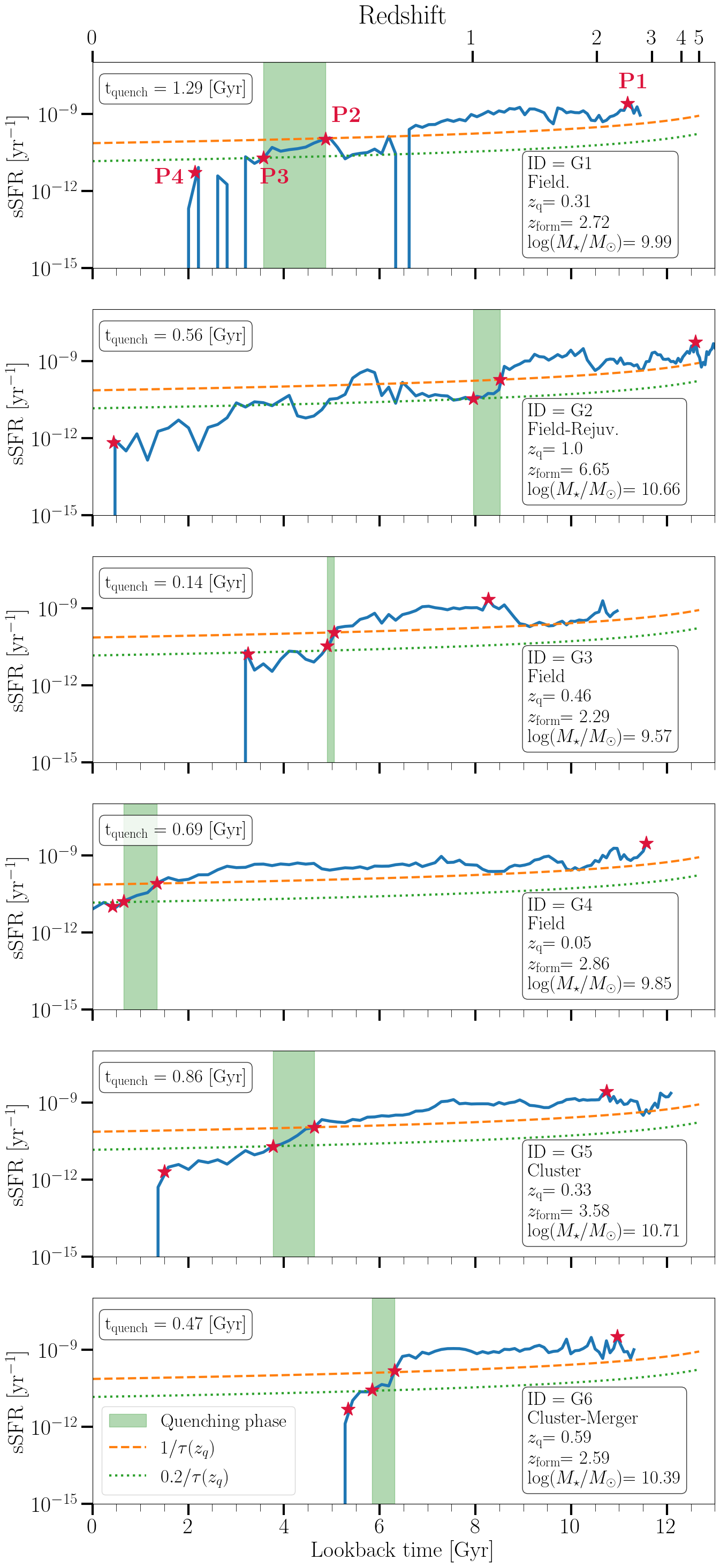}
\caption{Example sSFR versus look-back time for six representative QGs from this study. The sources' SFHs are illustrated with blue lines. Panels display QGs along with their corresponding names, quenching times ($\rm t_{\rm quench}$), the redshift of the quenching event ($z_{\rm q}$), the formation redshift ($z_{\rm form}$) and the stellar mass ($\log(M_{\star}/M_{\odot}$). Look-back time is shown on the bottom x-axis in Gyr, while the top x-axis shows the corresponding redshift. The sSFR is illustrated on the y-axis as log-scaled. The SFHs are superimposed with orange and green lines, which mark the thresholds for the start and end of the quenching event, respectively (see the description in Sect.~\ref{Identification of quenched galaxies}). The green vertical band delimits the duration of the quenching phase, while red stars are the four critical points we use to determine the sample evolution as described in Sect.~\ref{Evolution stages}.}
\label{fig:SFH_examples}
\end{figure}

We identify four main evolutionary points that we use in order to monitor the changes in ISM properties of our sample. These points can be understood as critical phases in the galaxy history since their formation, and are displayed in Fig.~\ref{fig:SFH_examples}. After the formation phase which is the first appearance of the galaxy in the simulation, we identify (Fig.~\ref{fig:SFH_examples}, from right to left, respectively) a peak phase (P1) which is the global peak of the galaxy star-formation history; quenching phase (P2) which is the start of the quenching process as identified in Sect.~\ref{Identification of quenched galaxies}; quenched phase (P3), which marks the end of the quenching phase; gas-removed phase (P4), the point at which the specific gas mass ($f_{\rm gas} = M_{\rm gas}/M_{\star}$) drops below the arbitrarily threshold of 0.1\% that we use to characterise the gas-dry stage. We adopt this limit as such amount of gas is practically negligible and close to the resolution limit of \texttt{SIMBA} for intermediate mass galaxies in our sample. This limit is connected to observations as it is an order of magnitude below the $f_{\rm gas}$ measured with ALMA for massive ($M_{\star} > 10^{11.5} M_{\odot}$) QGs at $z>1$ (\citealt{williams21}), QGs identified at $z\sim0.8$ in the LEGA-C survey (\citealt{spilker2018molecular}), and in local ellipticals (\citealt{2016Davis}). We are interested in studying how the ISM properties are linked to quenching timescales as well as to stellar properties such as age, stellar mass and distance to the MS. These properties can be measured from observations and can be used to predict the state of the ISM in a given QG. The use of the evolutionary points allows us to compare these global properties consistently for the whole sample. 

\subsection{Selection of cluster and field galaxies} \label{Cluster and field selection}

The analysis is performed using the full-physics SIMBA catalog, and contains 55609 galaxies at $z=0$. We select clusters using the dark halo mass at $z=0$, requiring a virial mass $M_{\rm halo} > 10^{14} M_{\odot}$. Clusters are then defined on the basis of their virial radius, which, despite not being representative of the entire structure, selects the region with the largest density contrast with respect to the field. We find a total of 25 clusters for a final sample of 3321 galaxies. 

Field galaxies are selected using a uniform random sampler on the \texttt{SIMBA} galaxies at $z=0$ that are outside the halos which we identify as clusters. The merger tree of each galaxy from the cluster sample has been reconstructed up to $z\sim2$ using the python library of CAESAR (\citealt{2014Thompson}), optimised for the handling of \texttt{SIMBA} data. Galaxies in the cluster sample at higher redshit may not be part of a virialized structure (\cite{chiang2013ancient}, \cite{lovell2018characterising}), but they will eventually be part of the dense environment we defined at $z=0$. This allows to ask if the evolution of QGs that end up in massive nad dense structures differ from purely field galaxies. The choice of not tracking directly the environment at each epoch avoid the ambiguity in defining high redshift large structures. For the field we used the public information available in the \texttt{SIMBA} catalogs repository.

We select the final sample of passive galaxies as described in Section ~\ref{Identification of quenched galaxies} obtaining a total of 2144 cluster QGs and 7436 field QGs. The galaxies for which it is possible to calculate a quenching time in the redshift range $0 \lesssim z \lesssim 2$ are selected for both cluster and field samples. Due to resolution effects, we limit the stellar mass range of the total \texttt{SIMBA} sample to $\log(M_{\star}/M_{\odot})>9$. In this way, we also ensure our sample to be mass-complete. 

\subsection{Identification of major and minor mergers}

We trace major and minor mergers along the galaxy history. Mergers are identified using the relative position of galaxies in the \texttt{SIMBA} catalogs. For each galaxy (primary) we explore a volume of $\sim \SI{0.5}{c\mega\parsec}$ in radius around it and trace the position of all other galaxies (secondary) in that region at each snapshot. We measure the maximum proper velocity of galaxies in clusters to be $\sim \SI{2000}{\kilo\meter\per\second}$ on average, and we approximate the typical timescale between snapshots in the redshift range $0 < z < 2$ as $\sim \SI{1e8}{\year}$, meaning that we assume the same galaxy to move of proper motion inside a volume of radius $\sim \SI{200}{c\kilo\parsec}$. We trace in this way each galaxy to its next position in the next snapshot. When a secondary galaxy enters a region of radius $\sim 10$ times the radius of the primary galaxy, it is classified as a potential merger. The secondary galaxy can then disappear, reappear or multiple smaller galaxies can temporarily form in that region due to the complex interaction and tidal forces. Each case is considered and evaluated in order to establish the validity of the merger. If the stellar mass ratio of the merging galaxies is $\rm R = M_{\rm secondary}/M_{primary} > 1:4$ then the merger is major merger, while if it is $\rm R < 1:4$ the merger is minor (\citealt{kaviraj2013significant}, \citealt{rodriguez2019mergers}). If multiple minor mergers are happening at the same time, their combined stellar mass ratio is compared to the major merger threshold $\rm R > 1:4$. Mergers exceeding this threshold are considered a single major merger undergoing intense morphological changes due to tidal forces.

\section{The global ISM evolution in cluster and field QGs} \label{Results}

To gain insight into the overall evolution of galaxy dust content in
our QGs, we use $f_{\rm dust} = M_{\rm dust}/M_{\star}$ and $\delta_{\rm DGR} = M_{\rm H_2}/M_{\star}$, which are well suited proxies to assess the efficiency of
the specific dust production and destruction mechanisms (\citealt{dunne2011herschel}, \citealt{rowlands2014dust}, \citealt{tan2014dust}, \citealt{bethermin2015evolution}, \citealt{calura2016dust},  \citealt{michalowski2019fate}, \citealt{burgarella2020observational}). A fundamental step in the evolution of these ISM properties is the quenching phase of the galaxy. The mechanism responsible for the quenching is considered to act on the molecular gas suppressing the SF. If the dust and molecular gas components are assumed to evolve together with the stellar component \citep{michalowski2019fate}, dust is also impacted by the quenching process. If thus the duration of the quenching process is assumed to be connected to its efficiency and, possibly, to entirely different mechanisms, studying $t_{\rm quench}$ will provide information on both the possible mechanism of quenching and its efficiency in affecting the ISM content.

\subsection{Evolution of the quenching mode with redshift} \label{Evolution of the quenching mode with redshift}

\begin{figure}[tbh!]
    \centering
    \includegraphics[width=0.92\columnwidth]{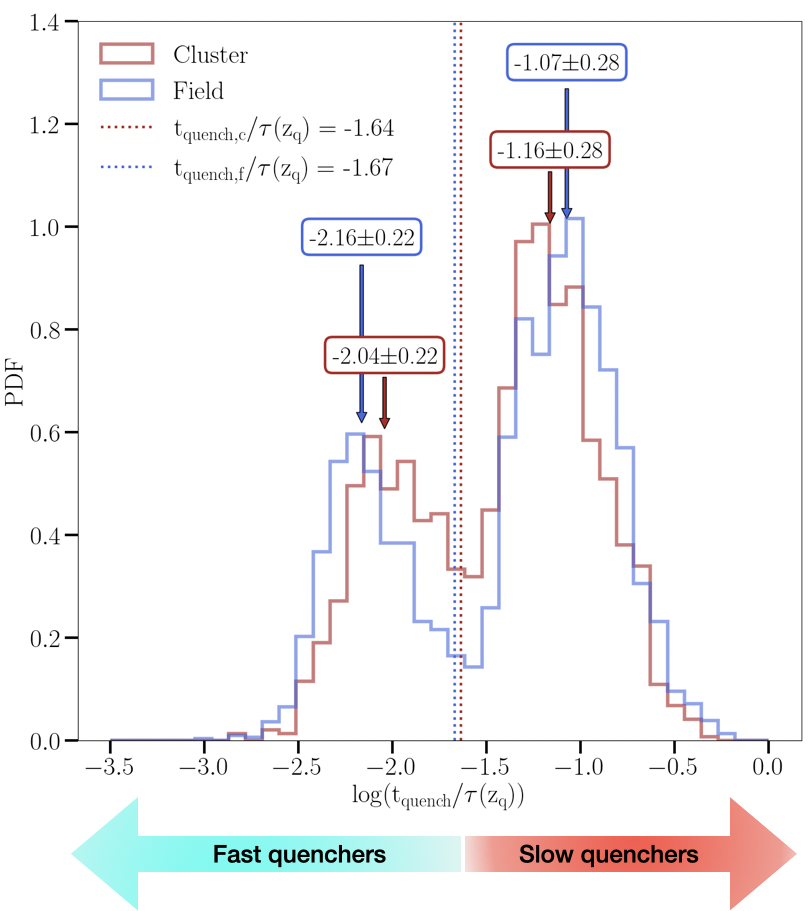}
    \caption{PDF distribution of the normalized quenching time for cluster (dark red) and field (blue) galaxies. The quenching time $\rm t_{\rm quench}$ is divided by the cosmic time $\tau(z_{\rm q})$ at the redshift $z_{\rm q}$ at which the galaxy starts quenching. The vertical dotted dark red(blue) line represents the $\log(\rm t_{\rm quench}/\tau(z_{\rm q})) = -1.64(-1.67)$ separation between fast and slow quenching galaxies in cluster(field).}
    \label{fig:QT_distr}
\end{figure}

Fig.~\ref{fig:QT_distr} shows the distribution of the quenching times as measured in Sect.~\ref{Identification of quenched galaxies}, normalized by the cosmic time $\tau(z_{\rm q})$ at the redshift of quenching $z_{\rm q}$. Both cluster (red) and field (blue) samples follows a clearly bimodal distribution. We perform a Gaussian fit for each of the two peaks in cluster (field), inferring  $t_{\rm quench, 1} = -2.04(-2.16) \pm 0.22(0.22)$ for the first peak and $t_{\rm quench, 2} = -1.16(-1.07) \pm 0.28(0.28)$ for the second, respectively. In terms of non-normalized timescales, the peak values correspond to $t_{\rm quench, 1} \sim 53.7(40.5) \rm Myr$ and $t_{\rm quench, 2} \sim 407.2(499.9) \rm Myr$ at $z \equiv 1$ and $t_{\rm quench, 1} \sim 30.1(22.74) \rm Myr$ and $t_{\rm quench, 2} \sim 228.4(280.5) \rm Myr$ at $z \equiv 2$ for cluster(field). We define two different regions of the distribution by calculating the intersection point of the Gaussian used to fit each peak. The obtained thresholds are $t_{\rm quench, c, lim}/\tau(z_{\rm q}) = -1.64$ and $t_{\rm quench, f, lim}/\tau(z_{\rm q}) = -1.67$ for cluster and field galaxies respectively. The bimodal distribution of normalised $t_{\rm quench}$ has been found in \texttt{SIMBA} for a similar range in stellar masses and redshift \citep{zheng2022rapidly} and up to $z \sim 4$ \citep{rodriguez2019mergers} with a similar position of the peaks and a separation threshold ($t_{\rm quench, lim}/\tau(z_{\rm q}) = -1.5$), roughly corresponding to $t_{\rm quench} \sim \SI{175}{\mega\year}$. 

Galaxies with associated $t_{\rm quench}/\tau(z_{\rm q})$ lower that the respective threshold are defined as fast QGs, while larger values identify what we call slow QGs. This definition consider slow QGs as galaxies that quench in a timescale comparable with the cosmic age at the time of quenching. The normalization by $\tau(z_{\rm q})$ ensures that this definition remain redshift-independent so that a galaxy that quench in a short timescale in the local Universe, can be considered slow quenching if its star formation ceases with the same timescale but in a much younger state of the Universe.

A comparison with the literature highlight how \texttt{SIMBA} predicts a peak of much shorter $t_{\rm quench}$ with respect to other simulations, with a separation threshold which is also generally smaller. \cite{Wright2019} found timescales of the order of some Gyr, with a division around $t_{\rm quench} \sim \SI{1.5}{\giga\year}$ from the \texttt{EAGLE} simulations. The characteristic $t_{\rm quench}$ dividing fast and slow QGs as obtained from the \texttt{IllustrisTNG} simulation is around $\rm t_{\rm quench} \sim \SI{1}{\giga\year}$ (\citealt{Walters22tng}). Furthermore, the duration of quenching events we infer in \texttt{SIMBA} qualitatively agrees with the fast and slow quenching timescales reported in observational studies of QGs at intermediate and high-$z$ (\citealt{socolovsky18}, \citealt{wu18}, \citealt{belli19},  \citealt{wu21}, \citealt{Tacchella_2022}, \citealt{hamadouche2023connection}, \citealt{park2023rapid}). Observations of massive QGs at $z\sim$0.8 showed a broad range of quenching timescales, indicative of a possible interplay of different quenching mechanisms in shaping the distribution of the sample (\citealt{Tacchella_2022}). A more environment-focused observational study found quenching timescales for clusters at $0.8 < z < 1.35$ to be around $\rm t_{\rm quench} = 1.0 \pm \SI{0.3}{\giga\year}$ (\citealt{foltz2018evolution}), which is consistent with the lower-end tail (near the separation threshold) of the slow quenching broad peak of the timescale distribution in \texttt{SIMBA} (see also \cite{zheng2022rapidly} Fig.2). These values, however, do not provide a clear categorization of the sample in fast or slow quenchers, rather inserting it into an undetermined category in our classification. In Sect.~\ref{discussion}, we provide comprehensive analysis on how the quenching routes and their underlying physical mechanisms affect the ISM gas and dust in QGs.

\begin{figure}[tbh!]
    \centering
    \includegraphics[width=0.85\columnwidth]{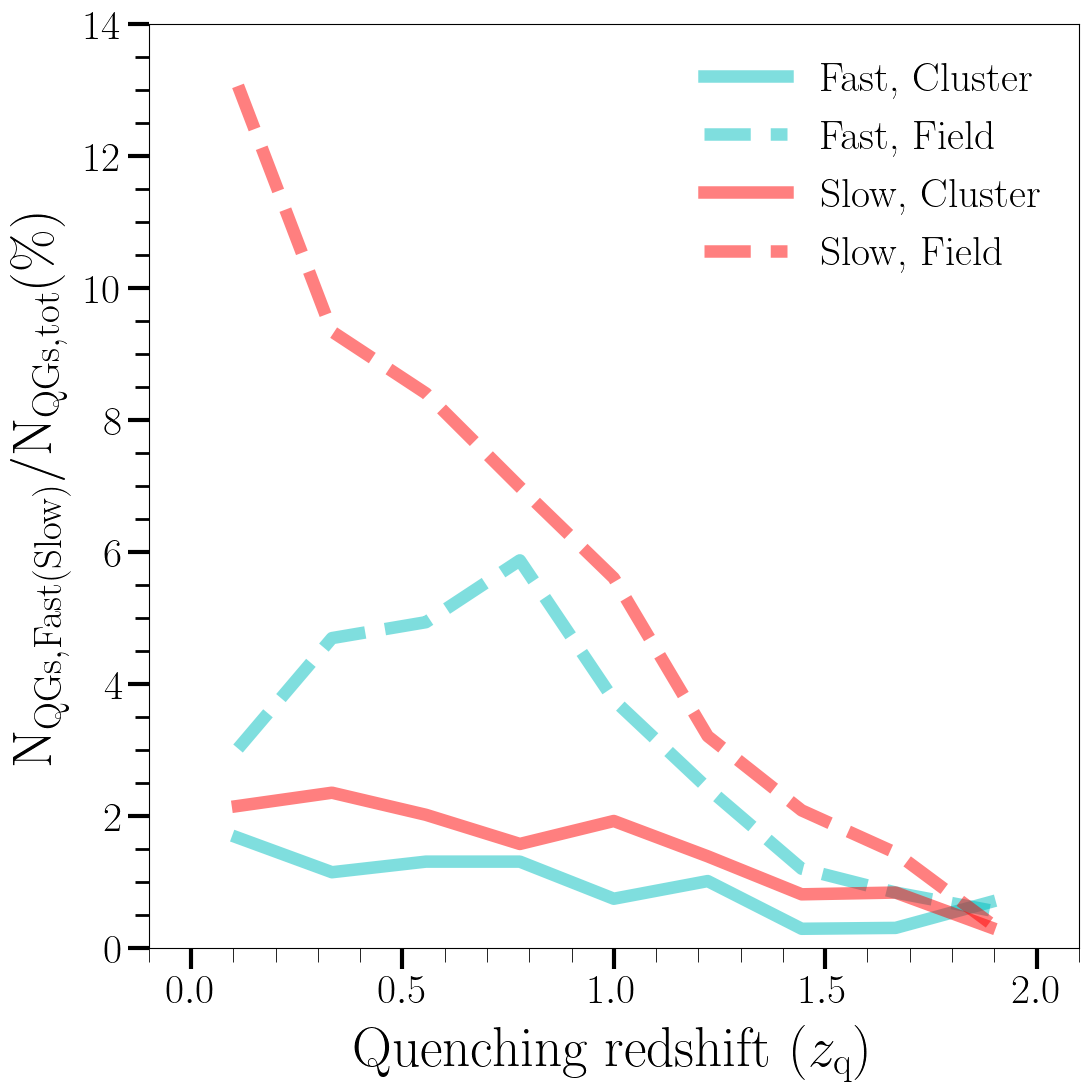}
    \caption{Fraction of fast (cyan) and slow (red) QGs with redshift for the cluster (continuous lines) and field (dashed lines).}
    \label{fig:qt_evol}
\end{figure}

To better have a sense of the evolution of $t_{\rm quench}$ with the cosmic time, we show in Fig.~\ref{fig:qt_evol} the fraction of fast and slow QGs at different redshifts with respect to the total number (cluster and field) of QGs in the sample. Normalising by the fixed size of the total sample allows to see how blindingly searching for QGs that quenched in the range $0 \lesssim z \lesssim 2$ would systematically lead to a larger probability of finding a QG in the field. The number of fast and slow quenchers for the cluster sample increase steadily towards $z=0$, with slow quenchers dominating at all redshifts. In the field the situation is analogous for galaxies that quench in the range $0.8 \lesssim z \lesssim 2.0$, but for $z \lesssim 0.8$ the number of fast quenchers rapidly decreases. This scenario is in line with both observations and semi-analytical model predictions of galaxies up to $z\sim3$ \citep{pandya2017nature} showing that at low-$z$ galaxies preferentially evolve through a slow quenching mode, while the fast quenchign mode increase in importance at higher redshifts. 

\subsection{The $\delta_{\rm DGR}$-metallicity plane for quiescent galaxies} \label{The dust-metallicity plane for quiescent galaxies}

Gas-phase metallicity is one of the crucial parameters influencing the dust life cycle in galaxies (e.g., \citealt{asano2013dust}, \citealt{wiseman17}, \citealt{popping2022observed}). The unique dust model implemented in \texttt{SIMBA} enables us to investigate how the different markers of galaxies' dust-life cycles (i.e., $\delta_{\rm DGR}$ and $f_{\rm dust}$) evolve as a function of $Z_{\rm gas}$ (expressed as $12+{\rm log(O/H)}$). 

\begin{figure*}[tbh!]
\centering
\includegraphics[width=\textwidth]{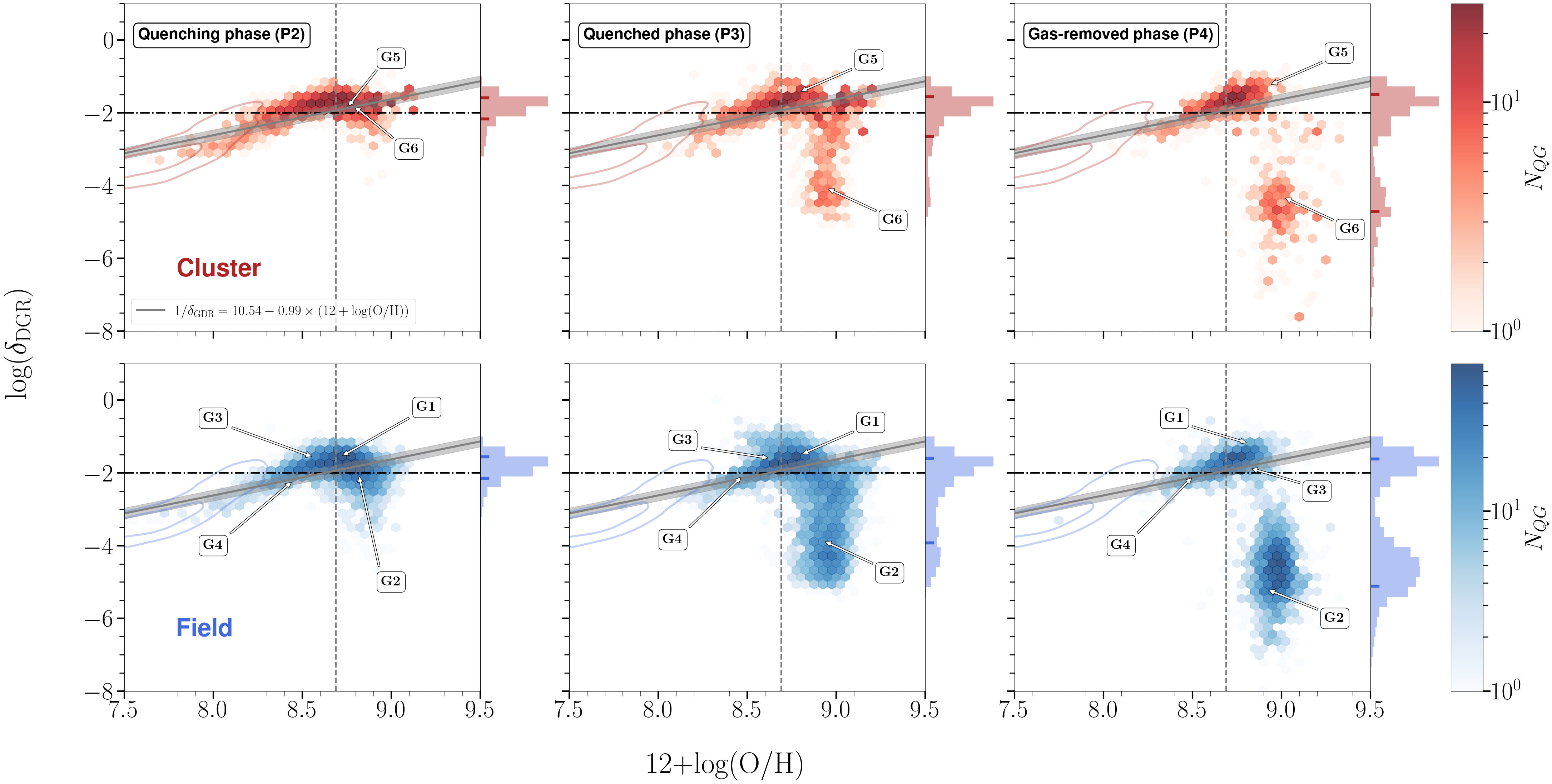}
\caption{The $\delta_{\rm DGR}$-metallicity plane for cluster QGs (upper panel) and field QGs (lower panel). Positions of QGs within the plane are shown for different evolutionary phases, as indicated in legend. Contours define positions typical for the peak star-formation phase. Colorbars refer to the number of sources. The vertical dashed line marks the solar value for the oxygen abundance ($Z_{\odot}$), while the horizontal dashed-dotted line shows the reference value for the dust-to-gas ratio in SFGs from the local Universe $\delta_{\rm DGR} = 1/100$. The positions of galaxies introduced in Fig.~\ref{fig:SFH_examples} are indicated with arrows. The scaling relation between $Z{\rm gas}$ and $\delta_{\rm DGR}$ for local SFGs (\citealt{magdis2012evolving}) is shown as a solid line with equation indicated in the legend. Histograms on the y-axis show the PDF distribution of $\log(\delta_{\rm DGR})$ for both samples, with small ticks marking the 16th and 84th percentile.}
\label{fig:DGR_plane}
\end{figure*}

In Fig.~\ref{fig:DGR_plane} we show the distribution on the dust-metallicity plane for cluster (field) QGs. The panels represent from top to bottom the samples observed at the start of their quenching phase (P2), at the end of the quenching phase (P3), and at the last evolutionary point (gas-removed phase, P4), respectively. Background contours show the 50th and 90th percentile level of the distribution at the peak of the SFH (P1). As described in Sect.~\ref{Evolution stages} we define the evolutionary point P4 using an arbitrary threshold for the cold gas fraction, namely $f_{H_2} = \rm M_{H_2}/\rm M_{\star} = 10^{-4}$. If a galaxy does not cross this threshold then it is not counted in the plane. The dashed vertical line marks the solar oxygen abundance $12+{\rm log(O/H)}=8.69$ \citep{asplund2009chemical}, while the horizontal dot-dashed line is the literature reference value ($\delta_{\rm DGR} \sim 1/100$, i.e., \citealt{lisenfeld2000dust}, \citealt{leroy2011co}, \citealt{magdis2011goods}, \citealt{sandstrom2013co}, \citealt{remy2014gas}, \citealt{whitaker21a}). Arrows show the position of the galaxies presented in Fig~\ref{fig:SFH_examples}. We show the empirical fit $\delta_{\rm DGR} = 10.54 - 0.99 \times (12+\log(\rm O/H)$ from \cite{magdis2012evolving} as a gray line, and we shade in gray the 0.15 dex scatter of the relation. The fit was done on a sample of star forming galaxies. 

Our sample recovers the same global trend observed for the entire population of galaxies at $z=0$ in \cite{simba2019}. The distribution has a characteristic, environment-independent "boomerang shape". It reconciles  a linear increase of the $\delta_{\rm DGR}$ with the oxygen abundance, combined with a large scatter towards super-solar $Z_{\rm gas}$ ($10^{-1} \leq \delta_{\rm DGR} \leq 10^{-5}$). 

Surprisingly, soon after quenching, $\sim$50\% ($\sim$40\%) of cluster (field) QGs attain $f_{\rm dust} \gtrsim 10^{-3}$ and thus roughly follow the trend expected for SF galaxies (i.e., best-fit relation from \cite{magdis2012evolving}). The scattered region which deviates from the fit is populated by $\sim$30\% (50\%) of cluster and field galaxies, respectively. As the galaxies evolve, the ISM reservoir sustains and has measurable $Z_{\rm gas}$, $M_{\rm dust}$ and $M_{\rm H_2}$, up to the quenched phase (P3) in $\sim95\%$ of QGs in the field sample. For cluster QGs, the fraction falls to $\sim 80\%$. In the remaining galaxies, the cold gas content is below the resolution of \texttt{SIMBA}, so that the properties of dust, metallicity and SFR are not measurable. The linear trend of the distribution is significantly populated even in the dry-phase (P4; $\sim$30\% for clusters and 20\% for flield QGs), showing no indication of an evolution with cosmic time. This behaviour resembles the one observed in DSFGs, for which redshift independent relation of $\delta_{\rm DGR}$ with $Z_{\rm gas}$ has been reported both at low redshifts ($z\sim0$, e.g., \citealt{de2019systematic}) and at high redshifts ($1.5 < z < 2.5$; e.g., \citealt{shapley2020first}, \citealt{popping2023dust}). 

The "boomerang-shaped" dust-metallicity plane from Fig.~\ref{fig:DGR_plane} reveals several interesting features: \textbf{(1)} The bulk of our QGs populate the same part of the plane as \texttt{SIMBA} SFGs (see Fig.3 in \citealt{li19}). Consequently, these QGs also follow the trend expected from the empirical relation between the $\delta_{\rm DGR}$ and $Z_{\rm gas}$ obtained for MS DSFGs (\citealt{magdis2012evolving}). Such well-behaved trend is mostly pronounced for galaxies of intermediate masses (see Appendix ~\ref{appendix1}). Intriguingly, a non-negligible number of QGs has $\delta_{\rm DGR}$ that is compatible or even higher, than observed for MS DSFGs. As an example, in Figure~\ref{fig:DGR_plane} we also highlight that four out of six QGs from Fig.~\ref{fig:SFH_examples} maintain (or even increase), their $\delta_{\rm DGR}$ during post-quenching evolution. 
\textbf{(2)} By confronting the upper and lower panels of Fig.~\ref{fig:DGR_plane} it is evident that the environment does not introduce the change in the shape of $\delta_{\rm DGR}$ and $Z_{\rm gas}$ plane, but reduces the dust abundance more efficiently in the cluster than in the field. Namely, both samples have a strong peak around the typical literature value of $\delta_{\rm DGR} \sim 1/100$, but field QGs cover the plane more uniformly along the region of large scatter. The difference between the distributions is strongest towards the last evolutionary point (P4) that describes the gas-dry phase. The fraction of QGs in this phase is rather similar, with only $45\%$($47\%$) of galaxies retaining a measurable amount of their ISM in the field and cluster environment, respectively. 

For $\sim10\%$ of the field QGs, gas fraction remains above our arbitrary threshold ($f_{\rm gas} > 10^{-4}$) over the entire evolution following the quenching event. Such QGs do not satisfy the "dry-phase" criteria, therefore their P4 point is not calculated. The actual fraction of field QGs with negligible trace of ISM is thus around 30-40\%. This number is only $\sim4\%$ for cluster QGs. For both cluster and field QGs the bulk of distribution remains centered around $\delta_{\rm DGR}\gtrsim 1/100$. These QGs tend to maintain a large fraction of dust during their last phases of passive evolution where cold gas content drops significantly; \textbf{(3)} At $12+\log(\rm O/H)\gtrsim8.8$ there is a range of $\delta_{\rm DGR}$ that extends over $\sim 4$ orders of magnitude independently of evolutionary phase or environment.
The absence of a clear trend in Fig.~\ref{fig:DGR_plane} suggests that the balance between formation and destruction of dust grains is strongly altered at super-solar $Z_{\rm gas}$. The large scatter appearing for metal-rich QGs has been observed locally in \citealp{casasola2020ism} and subsequently at intermediate redshifts in \citealp{donevski23}. A number of processes can contribute to this scatter, i.e., ISM-rich mergers, an increase in $f_{\rm gas}$ by environment-dependent inflows \citep{aoyama2022environmental}, and efficient re-growth of dust grains that can compensate the destruction due to thermal sputtering (\citealt{zhukovska2016modeling}, \citealt{pantoni2019new}, \citealt{Gillman2020}).
We explore possible scenarios in Sect.~\ref{discussion}. We note that throughout the rest of this work we show the analysis for the sample of field QGs since we clarified in Fig.~\ref{fig:DGR_plane} that the field and cluster samples have the same ISM properties even if differently distributed.

\subsection{ISM dust-gas abundance within the MS} \label{ISM dust-gas abundance within the MS}

\begin{figure*}[tbh!]
\centering
\resizebox{\hsize}{!}{\includegraphics[]{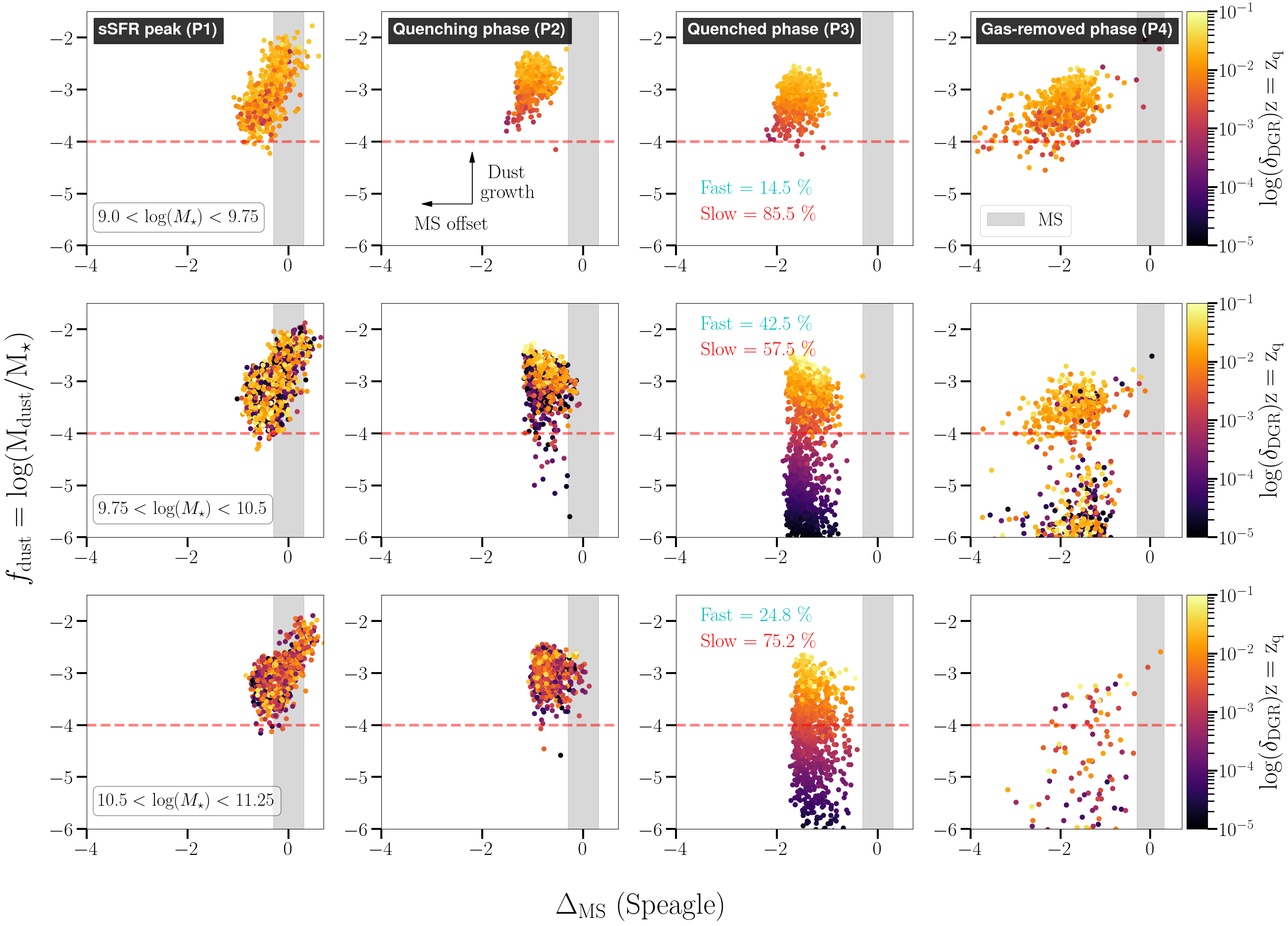}}
\caption{The evolution of $f_{\rm dust}$ with $\Delta_{\rm MS}$ for the field QGs. The $\Delta_{\rm MS}$ is derived following the best-fit relation from \cite{speagle2014highly}. Each column represents a different phase of the galaxy evolution as indicated in legend. These are, from left to right: sSFR-peak phase (P1), quenching phase (P2), quenched phase (P3), and gas-removed phase (P4). Each row corresponds to different stellar mass bins calculated at the quenching time. Stellar masses increase from top to bottom, while the sample evolves from left to right. The gray shaded region defines the 0.3 dex scatter of the MS. Points are color-coded for their $\delta_{\rm DGR}$ calculated at the quenched phase (P3). Points with the same colors in different columns represent the same galaxy across different evolutionary stages. The horizontal arrow marks the direction of evolution with respect to the MS, while the vertical arrow shows the direction expected for an increasing dust growth. For each panel, we show the fraction of fast(slow) QGs in the column corresponding to P3. To guide the eye, we show a dashed horizontal line ($f_{\rm dust} = 10^{-4}$), which roughly corresponds to cold gas fraction of $1\%$ and marks an approximate observational limit of known dusty QGs detected with ALMA (see Section~\ref{observational_studies}).
}
\label{fig:DGR_plane_evol}
\end{figure*}

To better understand the ISM evolution of our quenched sample, we trace the change in $f_{\rm dust}$ with respect to the offset from the MS ($\Delta_{\rm MS}$), as defined by \citealt{speagle2014highly}.\footnote{For this task, we homogenise the MS functional form from \cite{speagle2014highly} to the IMF by \cite{chabrier03}.}
In Fig.~\ref{fig:DGR_plane_evol} we illustrate this relation along the four key evolutionary stages introduced in Sect.~\ref{Evolution stages}.

As QGs evolve, they get further away from the MS, but we observe little to no changes in $f_{\rm dust}$ during the first two phases of their evolution. This behaviour is noticeable for all mass bins. Nevertheless, the later evolution of ISM dust and gas is different for QGs of different $M_{\star}$. Namely, for QGs of lower-masses ($\log(M_{\star}/M_{\odot})<9.75$) the dust content remains remarkably stable across entire evolutionary history. More complex behavior emerges for more massive QGs ($\log(M_{\star}/M_{\odot})\gtrsim9.75$), for which we observe a larger scatter in $f_{\rm dust}$ after the quenched phase. The QGs of these stellar masses show no clear evolution with $\Delta_{\rm MS}$. Our data indicate a general trend going from the top to the bottom row, with higher masses having on average lower $f_{\rm dust}$ at the gas-removed phase. This implies that $M_{\star}$ has important impact of the late evolution of ISM dust and gas in QGs, such that less massive QGs are characterised by a lower dust destruction rate and/or by a larger growth rate than in more massive galaxies. The significant spread in $f_{\rm dust}$ at the given MS-offset is interesting to explore, as QGs are generally expected to share similar, redshift independent SEDs, a consequence of a less complex and less turbulent ISM as compared to DSFGs (e.g., \citealt{magdis2021interstellar}). Therefore, it has been suggested that the relation between the $f_{\rm dust}$ and $\Delta_{\rm MS}$ can be interpreted as an "age-evolutionary sequence", such that QGs from the upper-right side of an $f_{\rm dust}-\Delta_{\rm MS}$ diagram are dominated by objects with younger stellar age and higher gas fraction. The later decrease in star formation rate (SFR) is likely due to exhaustion of their gas reservoirs, reflecting efficient gas removal (\citealt{gobat20}). However, the large scatter and absence of a clear trend during the later phases of evolution, suggest a complexity of processes affecting the ISM dust abundance. \cite{whitaker21a} suggested that a large range in $f_{\rm dust}$ hints at diverse evolutionary routes to quiescence, a question we thoroughly explore in the subsequent sections.

Furthermore, Fig.~\ref{fig:DGR_plane_evol} illustrates that $\delta_{\rm DGR}$ and $f_{\rm dust}$ follow each other until the start of the quenched phase. Hence we pay particular attention to the last column of the plot in Fig.~\ref{fig:DGR_plane_evol}. This evolutionary point is set around a specific value of $f_{H_2}$, that is $f_{H_2} \sim 10^{-4}$ (see Sect.~\ref{Evolution stages}), meaning that the main ISM property changing between the different galaxies inside a mass bin is $f_{\rm dust}$. At the start of the quenched phase, the scatter in $f_{\rm dust}$ presents a smooth gradient in $\delta_{\rm DGR}$, with low $\delta_{\rm DGR}$ being associated to low $f_{\rm dust}$. This trend is not maintained up to the final, gas-removed phase. The last column of Fig.~\ref{fig:DGR_plane_evol} showcases QGs with low $f_{\rm dust}$ and $\delta_{\rm DGR}$ in the quenched phase, attaining higher values of $f_{\rm dust}$ even with $f_{H_2} \sim 10^{-4}$. The vertical arrow indicates the direction along which the dust growth could have an effect. If there is a mechanism preventing dust grains to be destroyed efficiently, the major effect would be to increase both $f_{\rm dust}$ and $\delta_{\rm DGR}$. From the column corresponding to the P4 point in Fig.~\ref{fig:DGR_plane_evol} we see that the bulk of QGs with $f_{\rm dust} > 10^{-4}$ at the gas-removed phase is for low mass galaxies ($ 9 < \log(M_{\rm \star}/M_{\odot}) < 9.75$), but a non-negligible amount of galaxies in the range  $ 9.75 < \log(M_{\rm \star}/M_{\odot}) < 10.5$ and few galaxies are found for even larger masses. This suggests that even for QGs with virtually no cold ISM gas (that is the case in P4 by construction), copious amounts of dust can be found. The number of galaxies with large $f_{\rm dust}$ for the intermediate and large mass ranges is likely a consequence of the underling mass distribution as shown in the mass division of the $\delta_{\rm DGR}$-metallicity plane (Fig.~\ref{appendix1}).

\section{What are the mechanisms influencing the dust-gas evolution in high-$z$ QGs?} \label{discussion}  

To reveal the key sources influencing a complex dust-gas interplay presented above, in this Section we provide in-depth analysis of both external and internal mechanisms that may be responsible for the overall evolution of the ISM in QGs. The focus would be set in particular on exploring the origin of the large scatter in $f_{\rm dust}$.  This allows to characterize the properties of QGs with copious amount of dust in the ISM and to provide a qualitative evolution of their ISM content. 

\subsection{External channels affecting ISM dust evolution} \label{Mergers}

\subsubsection{Cluster environment}

As seen in Fig.~\ref{fig:DGR_plane}, the fraction of QGs with measurable ISM dust in the super-solar regime is reduced stronger for the cluster QGs than for the field ones. As a consequence, a large number of QGs with super-solar $Z_{\rm gas}$ has $\delta_{\rm DGR}$ that extends over >4 orders of magnitude, including exotically low values ($\delta_{\rm DGR}\lesssim10^{-4}$) that are first reported in \texttt{SIMBA} work by \cite{whitaker21a}. Such low values are similar to the low upper limit ($\delta_{\rm DGR} \sim 10^{-4}$) obtained from \textit{Planck} observations of galaxy clusters in the local Universe (\citealt{adam2016planck}). It is generally considered that thermal sputtering on dust grains embedded in a hot medium must be relatively rapid in massive halos typical for galaxy clusters (e.g., \citealt{galliano2021nearby}). The timescale for dust destruction by hot gas sputtering depends on the temperature $T_{\rm gas}$ and density $n_{\rm e}$ of the gas as $t_{\rm sp} = 10^{5}(1+(10^{6} \rm K/T_{\rm gas})^{3})n_{\rm e}^{-1} \rm yr$ (\citealp{hirashita2015dust}). This means that at the typical conditions of the hot halos of massive QGs ($T_{\rm gas} \sim \SI{1e7}{\kelvin}$, $n_{\rm e} \sim \SI{1e-3}{\cubic\centi\meter}$), the hot gas is expected to support rapid destruction of dust, with a timescale of the order of $t_{\rm sp} = \SI{1e8}{\year}$. Since the sputtering is efficient for $T_{\rm gas} > \SI{1e6}{\kelvin}$ and decreases rapidly with $T_{\rm gas} < \SI{1e6}{\kelvin}$ (\citealp{nozawa2006dust}), the effect is expected to be more significant in the cluster with respect to the field galaxies. The fraction of QGs in \texttt{SIMBA} during the dry, gas-removed phase is different for the two environments. We illustrate this in Appendix ~\ref{appendix1}, showing how the bulk of cluster QGs have almost entirely removed their ISM dust during the final evolutionary stage, in contrast to field QGs that keep the low, but measurable level of $f_{\rm dust}$. 

\cite{vogelsberger19} used the AREPO cosmological simulation (\citealt{springel2010pur}) and found a typical timescale of 1-10 Myr for the dust destruction within the cluster core. We calculate the timescale for \texttt{SIMBA} using the temperature of the circumgalactic medium (CGM), which in the simulation tops at around $0.5-1 \times 10^5$ K. For such temperatures the timescale is expected to be larger than for the typical scales of T$\sim 10^{6}-10^{7}$ K, namely $t_{\rm sp} \sim 100\:\rm Myr$. Despite this, a large population ($\sim$30\%) of galaxies in \texttt{SIMBA} clusters show dust fractions values well above the measurements of \cite{adam2016planck}, and consistent with MS galaxies (\citealp{magdis2012evolving}). This suggests a physical process that is able to alleviate the grain destruction due to thermal sputtering, most likely by prolonging the timescales by one order of magnitude at least. This is in line with recent works (\citealp{priestley2022properties}, \citealp{relano2022dust}) showing that longer sputtering timescales of $t_{\rm sp}>0.5-1$ Gyr support ongoing production of larger dust grains, which are expected to be reduced relatively slowly in hot halos. The possible dust re-growth scenario has also been proposed by \citealp{hirashita2015dust} and \citealp{hirashita2017dust}, as an explanation for the observed high $f_{\rm dust}$ in local elliptical QGs. These studies concluded that the variation of the dust growth efficiency is caused by large variations observed in $\delta_{\rm DGR}$, similar to those seen in our QGs. \cite{whitaker21a} conducted several numerical experiments with higher-resolution \texttt{SIMBA} runs and found that both thermal sputtering and dust growth are needed to explain the observed $\delta_{\rm DGR}$ in galaxies with low sSFR. We will return to this point in Sect.~\ref{sec:growth}.

\subsubsection{Mergers}

Merging events (in particular minor mergers) are important contributors to the size growth for massive QGs during their passive evolution (\cite{bundy2009greater}, \cite{kaviraj2013significant}, \cite{matharu2019hst}). Generally, mergers can form galaxies with different amount of dust depending on the dust content of the progenitors. 
Several studies propose that mergers with dust-rich galaxies can significantly enhance ISM dust abundance in QGs, potentially acting as the main channel for dust enrichment (\citealt{rowlands2012herschel}, \citealt{martini2013origin}, \citealt{lianou2016dustier}, \citealt{dariush2016h}, \citealt{kokusho2019dust}).  

We explore the occurrence of merging events among our QGs. From a complete track of the merger tree of QGs up to $z\sim2$ we infer that $\sim20\%$ of QGs experience at least one major merger in this redshift range and $\sim15\%$ of QGs experiencing multiple minor mergers. The major merger fraction is $\sim10\%$ during the gas-removal phase (from P2 to P4), which can last up to several Gyr and occurs mostly for $z<1$. This all confirms a non-negligible fraction of merging events in \texttt{SIMBA} galaxies as previously reported in \cite{rodriguez2019mergers}. Our merger fractions are consistent with those reported in observational study of  \cite{man2016resolving}. They use a large catalog of mergers observed up to $z\sim3$, and for major and minor mergers within the same redshift range as in this study, they infer $\sim20\%$ and $\sim15\%$, respectively. We further explore whether mergers are main drivers of the scatter in dust-related quantities as observed for the quenched phase in Fig.~\ref{fig:DGR_plane} and Fig.~\ref{fig:DGR_plane_evol}. We infer relatively low  fraction of mergers  with respect to the fraction of QGs populating the $\delta_{\rm DGR}$ scatter, which we measure to be $\sim45\%$($\sim20\%$) for QGs in the upper region of the scatter ($\log(\delta_{\rm DGR}) > -3$ and $12+\log(\rm O/H) > 8.69$), and $\sim55\%$($\sim36\%$) considering the entire scattered region (no constrain on $\delta_{\rm DGR}$) for P3(P4), respectively. We also check and confirm that the highest $\delta_{\rm DGR}$ and $f_{\rm dust}$ are not prevalent in mergers. Consequently, we conclude that observed amount of dust in QGs cannot be explained by considering the stochastic action of mergers only. The conclusion applies for both cluster and field QGs.   

Our conclusion is in line with recent observational results reporting that dust in QGs is not preferentially of external origin (e.g., \citealt{michalowski2019fate}, \citealt{blanquez2023gas}). It also qualitatively agrees with \cite{rodriguez2019mergers}. They trace the evolution of the merging and quenching rates in \texttt{SIMBA}, and report insufficient merger rate to explain the observed quenching rate at $z \leq 1.5$. Same study reported that fast quenchers are dominating the mass range around 1-3$\times 10^{10} M_{\odot}$, which corresponds to the second row of Fig.~\ref{fig:DGR_plane_evol}, where the scatter appears more prominent in both P3 and P4. Moreover, they do not find a significant physical connection between mergers and fast or slow quenchers. Since the scatter appears during the quenching phase, the conclusions of \cite{rodriguez2019mergers} suggest that a physical connection between the scatter and the merger process must be weak.  

\subsubsection{Rejuvenations}

As explored in \citealp{rodriguez2019mergers}, rejuvenation events can occur in \texttt{SIMBA} mostly in relation to minor mergers with gas-rich satellites. Bursts in the sSFR can be triggered by mergers and galaxy fly-bys and can thus be associated with a redistribution of the ISM in the galaxy and a burst in dust formation due to stellar activity. We make use of the empirical thresholds for star formation in order to identify rejuvenation events. We require a galaxy to be quenched for at least $0.2 \times \tau(z_{\rm q})$, with $\tau(z_{\rm q})$ being the cosmic time at quenching. A rejuvenation happens if the sSFR of a quenched galaxy is sSFR(z) $\geq$ 1/$\tau$(z), that is if the galaxy returns to the MS. The resulting fraction of rejuvenated galaxies is $\sim3\%$ for the clusters and $\sim2\%$ in field galaxies. Some examples of these objects are clearly seen in Fig.~\ref{fig:DGR_plane_evol}, as they re-renter the MS region during the late stages of evolution. We further check and find that our rejuvenated galaxies are not limited to a particular section of $\delta_{\rm DGR}$-metallicity plane, or $f_{\rm dust}$-MS plane, but populate its entirety. This all suggests that rejuvenations are not prevalent for producing a large scatter in $f_{\rm dust}$ and $\delta_{\rm DGR}$ in our QGs. 

Seemingly low rejuvenation fractions have recently been obtained from an independent work of \cite{szpila2024}. They use \texttt{Simba-C} (\citealt{hough2023simba}) QGs, and obtain a rapid downfall of rejuvenating fraction at $z<2$ to the point that $<5\%$ of galaxies at $z=0.5$ experience rejuvenation before $z=0$. The trend reverses at $z>2$, where \texttt{Simba-C} predicts a fraction $>30\%$, along with a population of high redshift QGs that reaches $z\sim5$. The low fraction of rejuvenations for low-$z$ galaxies can be misleadingly attributed to a lower amount of cosmic time left until $z=0$ for the rejuvenation to have a large chance of occurrence. However, observations of the contribution of O-type stars have shown that rejuvenation generally occurs during the first $<\SI{1}{\giga\year}$ after the quenching event (\citealt{zhang2023simple}). 

The lack of low-$z$ rejuvenated galaxies seems to be physically supported rather than an effect of redshift selection. Interestingly, some observations seem to point to an opposite trend, with the local Universe showing that 10-30\% of $z<0.1$ QGs have signs of ongoing rejuvenation (\citealt{treu2005assembly}, \citealt{schawinski2007effect}, \citealt{donas2007galex}, \citealt{thomas2010environment}), while this number decreases to an uncertain 1-16\% for $z>1$ (\citealt{belli2017kmos3d}, \citealt{chauke2019rejuvenation}, \citealt{Tacchella_2022}, \citealt{woodrum2022molecular}, \citealt{akhshik2021recent}). Such discrepancies are due to the lack of consensus on the definition of rejuvenated galaxy, of which an accurate comparison between simulated and observational data must carefully take into account, as highlighted by \cite{alarcon2023diffstar}. Here we explore a simple, but constraining definition for \texttt{SIMBA} rejuvenated galaxies, requesting the overall SF of a galaxy to be comparable with the low end of the MS, which can result in a selection of only the strongest and most extended onsets of new star formation. We also verify that lowering the SF threshold for the selection of rejuvenation to $\rm sSFR = 0.5/\tau(z)$ can increase the fraction to the observed 10-20\% at $z<2$. However this choice is not physically motivated. As described in \cite{szpila2024}, further constrains on the duration of the rejuvenation event would result in an even lower fraction. 

\subsection{Internal mechanisms}

Analysis from Sect.~\ref{Mergers} reveal that external mechanisms are not capable of producing the major effect on the observed dust-gas evolution in QGs. We therefore turn our attention to investigate in detail the internal processes affecting them. For simplicity, throughout the following sections, we present the trends for the field QGs only, but ensure that these do not change if we consider objects that reside in clusters.

\subsubsection{Impact of AGN feedback}

The presence of an AGN can have a significant impact on the evolution of the ISM. The energy from the AGN is injected into the ISM through the ejection of winds and high-energy radiation. Winds can sweep ISM material from the nuclear region of the galaxy into the CGM. If the accretion of matter into the central black hole is rapid enough, bipolar highly collimated jets may form and effectively remove the ISM material, increasing its temperature. Comparing the \texttt{Illustris}, \texttt{Illustris TNG} and \texttt{SIMBA} simulations, \cite{ma2022effects} concluded that the main effect of internal quenching has to be found in the cumulative energy released during the central black hole growth. This energy affects the formation of cold gas, leading to the cessation of star formation. Using an analogous diagnostic, we explore the effect of the AGN energy injection on the ISM.

\begin{figure*}[tbh!]
    \centering
    \includegraphics[width=\textwidth]{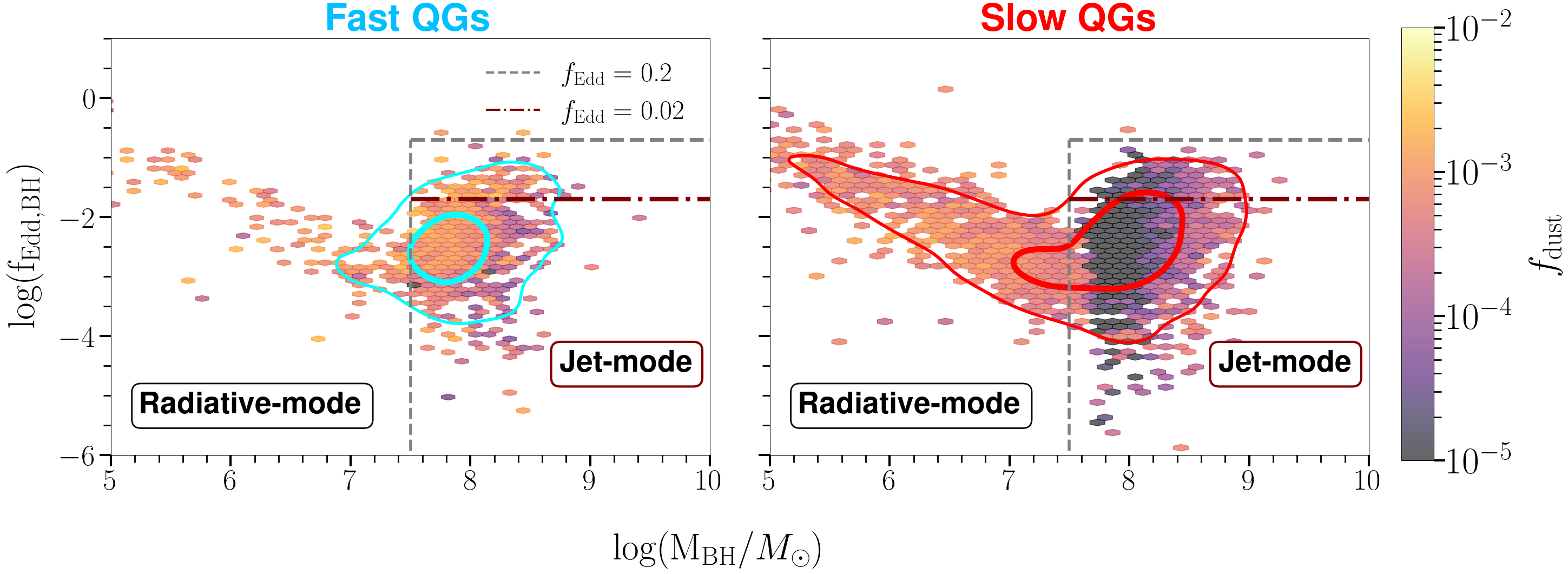}
    \caption{Distribution of the $f_{\rm Edd}$ over the central black hole mass $M_{\rm BH}$ of the sample of fast (left) and slow (right) field QGs at the quenched stage. The color coding shows the median dust fraction $f_{\rm dust}$ computed for the quenched stage (that is, at the P3 evolutionary point). The vertical dashed line represents the lower mass limit for the activation of jet mode in \texttt{SIMBA} ($\rm M_{\rm BH} = 10^{7.5}$). The dashed horizontal line delimits the starting of the transition from a radiative feedback mode to a jet mode ($f_{\rm Edd} = 0.2$), while the dashed-dotted line delimits the region with fully set jet mode ($f_{\rm Edd} < 0.02$). The red and blue contours represent the 50 and 90 percentile levels for the slow and fast quenching sub-samples respectively.}
    \label{fig:agn}
\end{figure*}

The jet mode of energy injection in the ISM from the AGN is triggered when $f_{\text{Edd}} < 0.02$ and $M_{\text{BH}} > 10^{7.5} M_{\odot}$ \cite{simba2019}. Figure~\ref{fig:agn} illustrates the effect of feedback variations (quantified with the Eddington ratio and $M_{\text{BH}}$) on the ISM dust abundance in QGs. We present the results for field QGs as a reference for comparison. All properties of the sample are measured at the quenched phase of the galaxies (P3), where the effect of star formation is minimal. The majority ($\sim90\%$) of fast-quenchers exclusively resides in the jet-dominated region of the plot. Slow quenchers are more evenly divided, with $\sim$26\% of QGs being influenced by the radiative-mode feedback, and jet-dominated AGNs influencing the rest. We also find that the AGN jet feedback is most commonly present in the stellar mass range of $1-5\times10^{10}M_{\odot}$, confirming the earlier results of \cite{rodriguez2019mergers}, \cite{appleby2020impact} and \cite{yang24}. 

A net drop of $\sim 2$ orders of magnitude in $f_{\rm dust}$ is visible going from the radiative-mode to the jet-mode dominated region in the sample of slow QGs. Moreover, QGs in the jet-dominated region of slow QGs have a larger scatter in their $f_{\rm dust}$ with respect to the QGs experiencing the radiative mode feedback, as visible in Fig.~\ref{fig:agn}. However, by looking at the impact of quenching routes on $f_{\rm dust}$ no explicit division can be found. Namely, the majority of both fast and slow quenchers appears dust-rich at the quenched phase, with median $\log(f_{\rm dust})\gtrsim-4$. Interestingly, fast quenchers have predominantly high $f_{\rm dust}$. Their median $\log(f_{\rm dust})$ is high in both feedback variants ($-3.09^{0.27}_{0.57}$ in the jet mode, and $-3.06^{0.25}_{0.39}$ in the radiative mode, respectively). These values are $2-5\times$ higher than those of slow quenchers ($-4.81^{1.18}_{0.83}$ and $-3.25^{0.26}_{0.45}$, respectively). In other words, while AGN jet feedback is primarily responsible for generating a population of rapidly quenched QGs, it does not correspond to an immediate removal of dust during the quenched phase.

This all points towards an important complexity between the timescales for cessation of star formation, gas removal and dust evolution in the late phases of QGs. The conclusion is qualitatively consistent with recent observational studies (e.g, \citealt{greene20}) suggesting that the peak of AGN activity may occur during an earlier phase of the quenching process but it is possible that the AGN jet-mode feedback has not fully cleared or destroyed the cold gas reservoir. Such a possibility is further sustained by \textit{Herschel} observations of large amount of dust in outflows from NGC 3079 in \cite{veilleux2021exploring}. They found that the sputtering timescale in the outflowing filament is consistent with the timescale of AGN activity required to lift the filament from the ISM. From the observation of star-formation-driven outflows in dwarf galaxies \cite{mccormick2018exploring} found dust entrained by galactic-scale winds, suggesting the possibility of shielded regions in which dust can survive to thermal sputtering and collisions. Dust surviving in the CGM was also observed by \cite{romano2024evidence}, likely carried out of the ISM by outflows. This provides a scenario in which the dust in expelled from the ISM, but not efficiently destroyed, and can then rain back down in the galaxy, as observed with the filament of NGC 3079 reconnecting to the galactic plane. The $H_2$ gas, on the other hand, is observed to be deposited in the CGM, where its destiny is uncertain and will depend on the AGN activity. This underlines possible different evolution pathways for dust and gas in the ISM. Further evidence comes from recent observations of dust-obscured quasars at $z\sim2.5$ showing a correlation between the reddening of the source and the presence of radiation-driven molecular outflows \citep{stacey2022red}. The study found that dust and molecular gas can survive after being removed from the galaxy even in the most luminous quasar in a process that effectively and rapidly quench the galaxy nuclear region. This scenario has been supported by high resolution simulations of galactic winds as \texttt{AREPO-RT} \citep{kannan2021dust} and \texttt{Cholla} (\citealt{schneider2015cholla}) for different grain sizes and distances from the galactic plane \citep{richie2024dust}. Dust survival will depend on the conditions of the ISM, with SNe-heated gas being more efficient in destroying dust than gas affected by the AGN. After being being deposited on the CGM, the surviving material can eventually rain down on the galaxy bringing new gas and dust to the ISM. The amount of ISM replenished in this way would depend on the potential well of the galaxy, the energy injected to the outflowing material, and the conditions of the CGM. This delicate balance is not yet fully understood. In Sect.~\ref{sec:timescale}, we further investigate this possibility by analysing the properties of fast and slow QGs and their influence on the $f_{\rm dust}$ evolution. 

\subsubsection{X-ray feedback}

AGNs in \texttt{SIMBA} contribute to the energy injection into the ISM and CGM also through X-ray radiation. This effect is triggered during the full-velocity jet-phase of the AGN and it depends on $f_{\rm gas}$, since the gas surrounding SMBHs can absorb and radiate away the X-ray energy. The main effect of the X-ray contribution is on the heating of the gas component, on the non-ISM gas and a combination of heating and momentum transfer for the ISM gas around the SMBH. This leads to two main consequences: on the one hand, massive halos have larger $T_{\rm gas}$, especially in the central region, as observed in X-ray profiles of \texttt{SIMBA} by \cite{robson2020x};  on the other hand, gas is efficiently removed from low-mass halos and is pushed outwards from the central region of galaxies, preventing formation of new stars, as found in \cite{appleby2020impact}. 

At the start of the quenching phase, $\sim$30\% of both field and cluster QGs from our sample satisfy the condition for X-ray heating activation. This fraction, however, increases differently towards the end of a quenching phase, reaching $\sim$60\% for the field QGs, and only $\sim$40\% for the cluster QGs. This is likely connected to a slightly larger fraction of radiative-mode dominated QGs in clusters, since galaxies that do not experience jet-mode do not trigger X-ray heating in \texttt{SIMBA}. The lack of strong AGN feedback manifesting through both collimated jets and powerful X-ray radiation can help explain the extended tail of slowly-quenched QGs containing substantial ISM dust abundance. Such QGs are not observed amongst the fast quenchers.

\subsubsection{Impact of quenching routes and timescales of $\rm H_{2}$ gas removal on the dust content of QGs} \label{sec:timescale}

One of the main questions arising from the presented results is how the dust in QGs is impacted by the different quenching routes and the consequent variety of timescales for the cold gas removal that follows the quenching process. The quenching timescale (as introduced in Sect.~\ref{Evolution stages}) reflects the transition period leading a galaxy to quiescence, and it is linked to the quenching mechanism itself. If, after the cessation of the SF, the changes of cold gas and dust tightly follow each other, one can conclude that the evolution in $f_{\rm dust}$ is solely dependent on the quenching route (see e.g., \citealt{gobat20}). This assumption is at the basis of the age-evolutionary framework, which proposes that once quenched, QGs undergo consumption (or destruction) of the dust and cold gas without a further replenishment. Such evolutionary picture requires a strong anticorrelation of $f_{\rm dust}$ with stellar age (\citealt{michalowski2019fate}, \citealt{caliendo21}, \citealt{whitaker21a}, \citealt{blanquez2023gas}). 

\begin{figure*}[tbh!]
	\centering
	\includegraphics[width=0.97\textwidth]{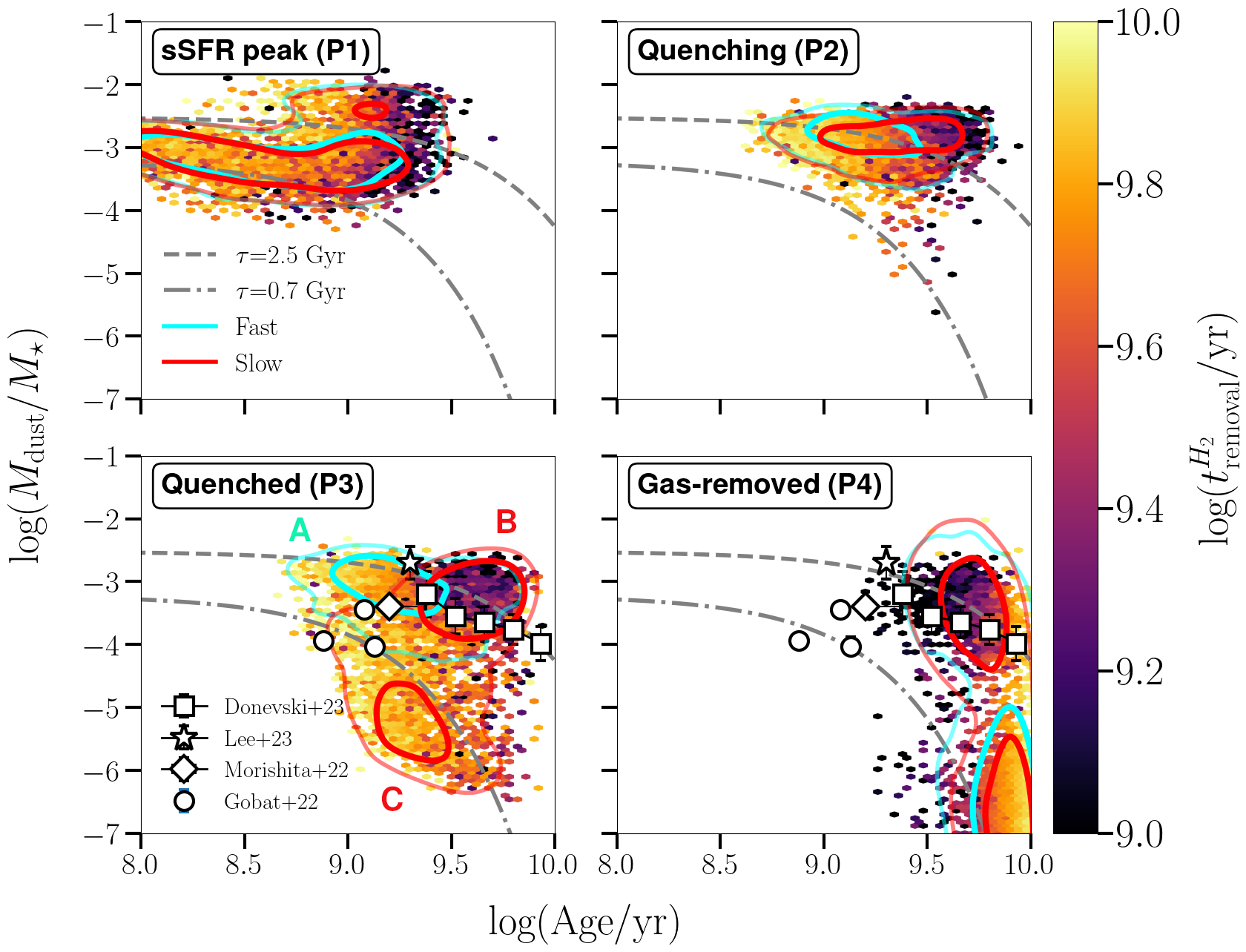}
	\caption{Evolution of $f_{\rm dust}$ with stellar age. We track the same evolutionary stages as introduced in Sect.~\ref{ISM dust-gas abundance within the MS}. Points are colour-coded with the mean quenched timescale, that is the time interval between the P2 and P4. Gray dashed and dashed-dotted lines show the exponential decrease of dust abundance assuming a timescale of $\tau = \SI{2.5}{\giga\year}$ and $\tau = \SI{0.7}{\giga\year}$ respectively, and are motivated by \cite{michalowski2019fate} and \cite{lee2023high}. White squares denote the binned medians from the sample of dusty QGs observed in the COSMOS field (\citealp{donevski23}). Other markers show various QGs followed-up with ALMA: empty star ($z=2$, \citealp{lee2023high}), empty circles ($z\sim1.5-2$, \citealt{gobat2022uncertain}), and empty diamond ($z=2.149$, \citealt{morishita22}). Cyan (red) contours define the 50th and 90th percentile of the distribution inferred for the fast (slow) QGs from this study, respectively. In the bottom left panel the three main peaks of the distribution are labeled A, B, and C.}
	\label{fig:f_dust_age_exponential_evol}
\end{figure*}

To  probe this scenario, we explore the overall evolution of $f_{\rm dust}$ with the mass-weighted stellar age. Figure~\ref{fig:f_dust_age_exponential_evol} showcases this relation for the same four critical points of SFHs as presented in Fig.~\ref{fig:DGR_plane_evol}. We colour-code with the timescales for cold gas removal, defined as time elapsing between the beginning of a quenching phase (P2) and the beginning of the subsequent gas-removed phase (P4) as:
\begin{equation}
	t^{H_2}_{\rm removal} = \tau(z_{P4}) - \tau(z_{P2}),
\end{equation}

where $\tau(z)$ is the cosmic time at z. Cyan(red) contours represents the 50th and 90th percentile of the distribution of the fast(slow) QGs. Fig.~\ref{fig:f_dust_age_exponential_evol} reveals that dusty QGs in \texttt{SIMBA} start their quenching phase with a wide distribution in stellar ages and with a relatively narrow range of $f_{\rm dust}$. While moving from the peak of sSFR to the start of the quenching event, stellar ageing in QGs does not introduce significant scatter in $f_{\rm dust}$. The situation drastically changes when QGs transition to the quenched stage. At the given stellar age, this evolutionary point unveils not only a significant scatter in $f_{\rm dust}$, but also a division between the underlying quenching routes.

The lower left panel of Fig.~\ref{fig:f_dust_age_exponential_evol} reveals three prominent regions. The first one is reserved for fast quenchers that occupy upper left side of the plane, encompassing dustier objects with younger stellar populations (A). This region only partly overlaps with those of slow quenchers, who are characterised with generally older stellar populations ($\log(\rm Age/yr)\gtrsim9.4$). Slow quenchers consist of two distinct clouds: one of slow, relatively dust-rich QGs (B; $\log(f_{\rm dust})\gtrsim -4$), and one of slow dust-poor QGs (C; $\log(f_{\rm dust})\lesssim -5$). The positions of these three clouds change when QGs reach their final phase, characterised with significant deficit of cold $H_{\rm 2}$ gas. While both group A and B end-up dust-poor and populate the lower-right side of the diagram, group C maintain an extremely shallow or almost absent evolution in $f_{\rm dust}$, contrasting the expectations from age-evolutionary models.

To understand the diverse pathways for dust evolution amongst the three distinct QG groups mentioned above, we first check their median quenching times and find them to be $t_{\rm quench, A} \equiv 0.05^{+0.05}_{-0.02}$ Gyr for fast quenchers, and $t_{\rm quench, B} \equiv 0.82^{+0.98}_{-0.48}$ Gyr and $t_{\rm quench, C} \equiv 0.61^{+0.70}_{-0.28}$ Gyr, for dust-rich and dust-poor slow quenchers, respectively. Values are presented in Tab~\ref{tab:examples_values} together with other characterizing properties of the sub-samples. Errors are assigned as the 16th and 84th percentile of each sub-sample distribution. The scale of the error is a direct consequence of the fast and slow QGs definition, which is not directly based on $t_{\rm quench}$, but accounts for a normalization by the cosmic time at the epoch of quenching as explained in Sect.~\ref{Evolution of the quenching mode with redshift}. This means that slow QGs can have both short and long $t_{\rm quench}$ if they quenched at low-$z$ or high-$z$ respectively, spreading out the distribution.  

If we focus on the median of the $t_{\rm quench}$ it appears that the division between fast and slow QGs is much clearer than a division between the groups B and C that compose the slow quenchers. Therefore, migration of the slow QGs to low $f_{\rm dust}$ does not depend solely on the duration of the quenching event, but is largely affected by the dominating quenching mechanism. Both the A and C groups are mostly affected by a jet-dominated AGN, while the B group is influenced by a radiative-mode feedback, as shown in Fig.~\ref{fig:agn}.

\cite{akins2022quenching} coupled \texttt{SIMBA} with \texttt{POWDERDAY} radiative transfer code \citep{narayanan2021powderday}, and tracked the main evolutionary routes among galaxies transitioning through the quenching phase and those that are recently quenched after an intense episode of starburst (so-called post-starbursts galaxies, PSBs). Interestingly, they find further movement across the UVJ colour plane independent of quenching timescales. As a result the quenching mode itself seems to be insufficient to characterize the ISM content of QGs that quenched at $z\sim1-2$ (see also \citealt{sun2022post}, \citealt{french2023state}). The C group from our sample show analogous characteristics to the PSBs described. Such galaxies have slightly higher $f_{\rm H_2}$ with respect to other subsamples, $\log(f_{\rm H_2, B}) = -1.37^{+0.17}_{-0.17}$, $\log(f_{\rm H_2, A}) =-1.37^{+0.27}_{-0.55}$, and $\log(f_{\rm H_2, C}) = -1.20^{+0.13}_{-0.14}$ respectively. 

Higher incidence rate of AGN feedback in QGs (and/or PSB galaxies) with younger stellar ages has recently been confirmed by several studies (\citealt{greene20}, \citealt{wu23}). Additionally, \cite{lammers23} explored MaNGA \citep{bundy2014overview} galaxies, and found that the radio-mode AGN feedback that significantly affects central SF may not always be efficient in expelling cold gas in a galaxy-wide fashion. They infer a long-lasting timescale for gas removal of $\sim5-6$ Gyr. The cold gas removal timescales obtained for rapidly quenched dusty QGs in this study are similarly long (see Tab.~\ref{tab:examples_values}). This may support the idea that even for rapidly-quenched QGs that are being affected by jet-mode feedback, AGN efficiency is not directly translated to the further dust evolution.

\subsubsection{Dissecting the age-evolution of dusty QGs}
\label{observational_studies}

The overall evolution and the large scatter observed in $f_{\rm dust}$ at the quenched phase is not resolved by considering the mass weighted stellar age as presented in Fig.~\ref{fig:f_dust_age_exponential_evol}. Within the age-evolutionary scenario, both the dust and cold gas content of QGs are expected to follow each other and decline exponentially as stellar population ages, as proposed in several studies (e.g., \citealt{michalowski2019fate}, \citealt{li19},  \citealt{nanni20}, \citealt{lee2023high}). This evolution takes the functional form:
\begin{equation}
	f_{\rm dust} = \frac{M_{\text{dust}}}{M_{\star}} \simeq A \cdot {\rm exp} \left(-\frac{\rm Age}{\tau}\right),
\end{equation}

where A is a normalization constant and $\tau$ is the \textit{e}-folding time. The typical timescale of dust removal is still uncertain, with some observations proposing values of $\tau > 1$ Gyr (\citealt{michalowski2019fate}, \citealt{gobat2018unexpectedly}, \citealt{lee2023high}) and others stating much faster timescales $\tau < 100 -500$ Myr (\citealt{williams21}, \citealt{whitaker21b}). For comparison, we show in Fig.~\ref{fig:f_dust_age_exponential_evol} the evolution of $f_{\rm dust}$ with age adopting a timescale of $\tau \equiv \SI{0.7}{\giga\year}$ and one of $\tau \equiv \SI{2.5}{\giga\year}$. While part of the distribution of \texttt{SIMBA} QGs falls into the region spanned by the expected timescales, a cloud of old QGs (log(${\rm age}_{\star}$/yr) $\gtrsim$ 9.4) maintains relatively large dust content ($-3<\log f_{\rm dust})<-4$). These values are $3-5\times$ larger than expected from the longest dust destruction timescale proposed by \cite{michalowski2019fate}. On top of this, the evolution of QGs that become dust-poor ($\log(f_{\rm dust})\lesssim-5$) after the quenching phase, appear to be much steeper than empirically predicted. These galaxies move almost vertically on the plane, from the quenching phase to the quenched phase, in the time $t_{\rm quench}$. Contours show the sample distribution for the fast and slow QGs, highlighting how the scatter generated during the quenching phase is intrinsic to the slow QGs. The evolution along the dust-age plane for the specific examples galaxies of Fig.~\ref{fig:SFH_examples} are shown and discussed in Appendix~\ref{appendix2}.

The fact that the $\rm H_2$ removal timescale can be short even for relatively dusty QGs, challenges the common idea that the dust and $\rm H_2$ components of the ISM evolve alike. From Tab~\ref{tab:examples_values} we see that mean values of $f_{\rm H_2}$ are similar for all three sub-samples at the quenched stage, while the dust amount in QGs from group C is systematically reduced, resulting in average $\delta_{DGR}$ and $f_{\rm dust}$ being >2 orders of magnitude lower than in other QGs. This implies that caution should be taken when inferring $M_{\rm rm gas}$ from $M_{\rm dust}$, as highlighted in \cite{whitaker21a}. 

\begin{figure*}[tbh!]
	\centering
	\includegraphics[width=0.69\textwidth]{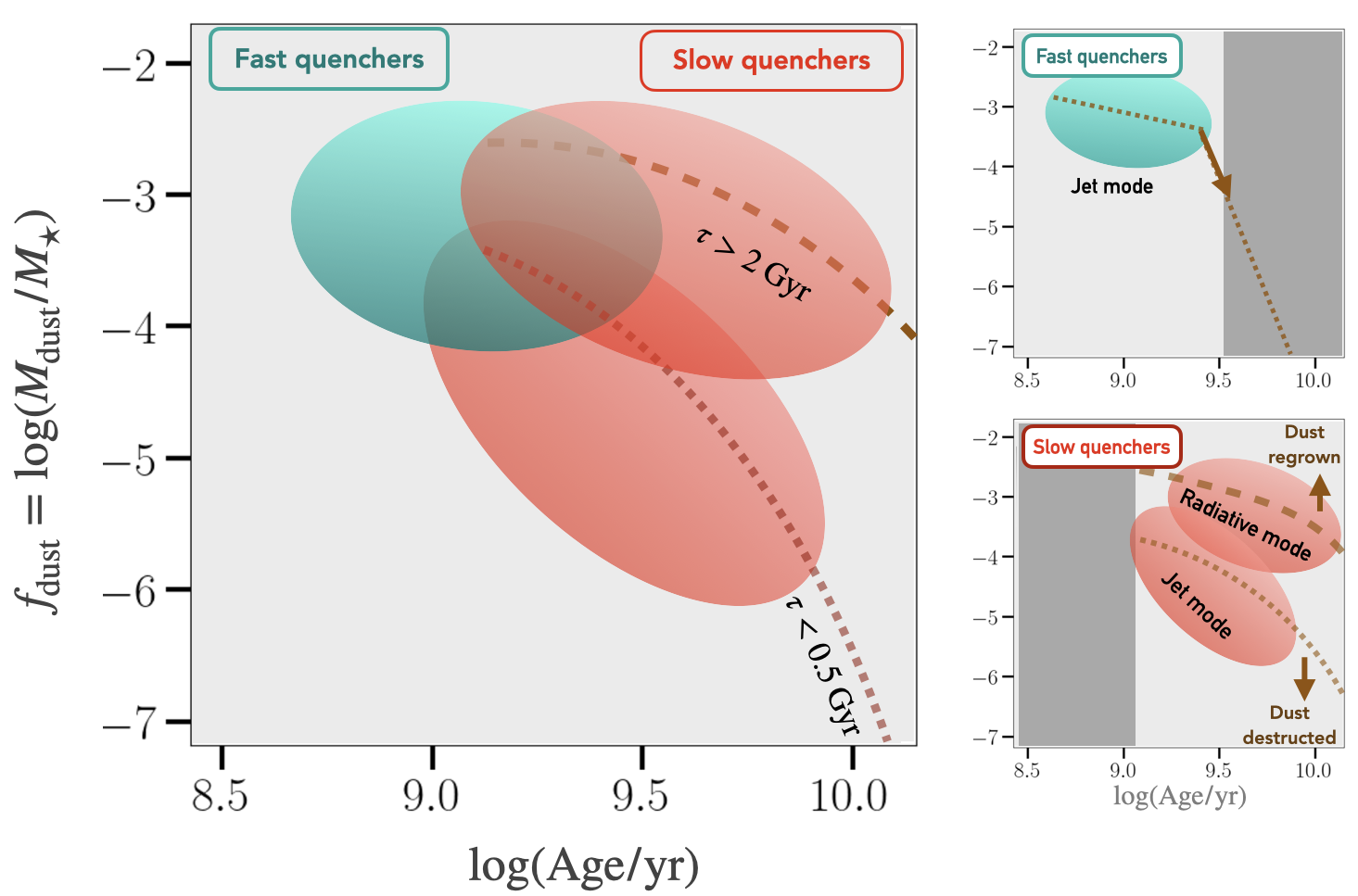}
	\caption{A qualitative sketch explaining the evolution of $f_{\rm dust}$ with stellar age within the quenched phase, as presented in the bottom panel of Fig.~\ref{fig:f_dust_age_exponential_evol}. Cyan cloud illustrates the position of fast quenchers, while two red clouds illustrate the position of dust-poor and dust-rich slow quenchers. The evolutionary directions (marked with arrows) for each sample are shown along with the main physical mechanisms involved in their dust evolution. The meaning of lines in the large panel is the same as in Fig.~\ref{fig:f_dust_age_exponential_evol}. The dark-grey shaded regions in the two subplots from the right panel delineate the part of the age-dust plane devoid of the given QG.}
    \label{fig:scketch}
\end{figure*}

Finally, we compare the distribution of simulated galaxies with observations from \cite{morishita22}, \cite{lee2023high}, \cite{gobat2022uncertain}, and \cite{donevski23}. The QGs from the first three studies reside at $z\sim1.5-2$ while QGs from the later study are at intermediate redshifts of $0.2<z<0.7$. The spread of the observational points agrees with the large scatter in $f_{\rm dust}$ predicted by our \texttt{SIMBA} analysis. From the two lower panels in Fig. ~\ref{fig:f_dust_age_exponential_evol} we can reveal the possible nature of $f_{\rm dust}$ as well as evolutionary stage of observed QGs. Individually detected QGs with younger stellar ages are caught after the completion of their quenching phase, and are mostly consistent with the fast mode of quenching. The particularly prominent object from Fig.~\ref{fig:f_dust_age_exponential_evol} is the "dustiest" QG identified at $z\sim2$ by \cite{lee2023high}. The position of this source within the $f_{\rm dust}$-age plane coincides with group A QGs which suggests the role of AGN jet feedback. Despite this, the $f_{\rm dust}$ is still considerably high due to a brief and efficient episode of dust growth, while the cold gas still preserves detectable amount over $\sim \rm Gyr$ timescales. This interpretation matches remarkably well the observational characterisation inferred by \cite{lee2023high}, that is, evidence for a fast quenching episode but fairly slow ISM removal (see also \citealt{kakimoto24}). The same authors pointed out that the excess of dust observed for their source is difficult to reproduce assuming stellar dust production, even when accounting for maximal SNe yields with no destruction. This implies a support of some additional dust re-formation processes. Switching to other observations of QGs with similarly young stellar populations we can notice substantially lower $f_{\rm dust}$ as compared to the QG from \cite{lee2023high}. The QG from \cite{morishita22} has a dust temperature ($\sim 28 \rm K$) warmer than in typical QGs. This is a consequence of an extremely strong AGN which is found to power the galaxy-scale outflows yet leaving a substantial content of warm gas long after quenching (\citealt{man21}). Such conditions may suppress the efficient growth which may further explain the drop in $f_{\rm dust}$. Interestingly, the two most massive QGs ($M_{\star}>10^{11}M_{\odot}$) have the lowest $f_{\rm dust}$ and come from the sample of strongly lensed QGs analysed by \cite{gobat2022uncertain}. They are on the track consistent with the fast dust depletion, in agreement with the suggestion from \cite{whitaker21b}. Nevertheless, caution should be taken for these objects, as their $M_{\rm dust}$ may be subject to larger uncertainties due to their extended sizes, as discussed in \cite{gobat2022uncertain}. 

The majority of QGs from \cite{donevski23} occupy the region dominated by slow quenchers, but are not exclusively attributed to the quenched phase. As can be seen from the bottom-right panel of Fig.~\ref{fig:f_dust_age_exponential_evol}, the oldest QGs may be even entering the dry-phase devoid of significant cold gas. The expected gas-removal timescale of $t^{H_2}_{\rm removal} \lesssim 1 \rm Gyr$ predicted by \texttt{SIMBA} is in agreement with what \cite{donevski23} derived for early-type QGs as the amount of time elapsed between the quenching and the time of epoch of the observation. Their sample has been selected excluding the presence of X-ray bright AGN candidates, which supports the dominance of low-efficiency radiative feedback as seen for the galaxies in group B in this study. 

\subsection{Emergence of dusty QGs via prolonged dust growth: a new pathway for the late ISM evolution} \label{sec:growth}

The analyses from previous sections reveal the existence of a considerable number of QGs for which the dust evolution following the quenching event occurs independently of gas-removal timescales. The most striking examples are dust-rich QGs for which $f_{\rm dust}$ remains high even in a late phase characterised with a cold gas deficit. 
Following the Eq.2, we infer the median values for the gas-removal timescales, and find them to be
$t^{H_2}_{\rm removal, A} = 5.73^{+2.18}_{-3.00}$ Gyr, $t^{H_2}_{\rm removal, B} = 2.81^{+4.22}_{-1.66}$ Gyr, $t^{H_2}_{\rm removal, C} = 6.42^{+1.80}_{-2.54}$ Gyr. From this we can deduce that in dusty QGs the gas-removal timescales are not tightly related to the dominant quenching mode, introducing new complexity in the pathways for the dust and gas (co-)evolution. 

In QGs dominated by older stellar populations, there would be time for AGB stars to contribute to the dust mass budget. However, as pointed out in several works (\citealt{nanni13}, \citealt{hirashita2015dust}, \citealt{ventura20}), the dust mass production efficiency decreases in metal-rich environments similar to our QGs. On top of this, the AGB condensation efficiency adopted in \texttt{SIMBA} models is quite low ($<20\%$), which would return the maximal $M_{\rm dust}$ order of $\lesssim10^{6}M_{\odot}$, unable to capture the higher values seen in our dusty QGs. This suggests an insufficient dust yield solely produced by AGB stars which prevents them for being a \textit{dominant} channel of dust re-formation in QGs, in line with conclusions from several independent works (e.g, \citealt{hirashita2015dust}, \citealt{michalowski2019fate}).

To explain the excess of dust content of QGs, a handful of studies pointed to the importance of dust growth that would allow the dust grains to increase in size, thus counteracting the joint actions of AGN and thermal sputtering (\citealt{mattsson14}, \citealt{hirashita2017dust}, \citealt{richings18}, \citealt{whitaker21a}, \citealt{donevski23}). This process is expected to be viable in metal-rich environments once the heated gas cools down. To probe this scenario in a quantitative way, we first check a dust re-growth timescale ($\tau_{\rm acc}$) in QGs which we roughly derive following \cite{liqi19} and using the same prescription used for \texttt{SIMBA}:
\begin{equation}
\begin{aligned}
\tau_{\rm acc} = & \, 1\times 10^{7} \, \text{yr} \\
                 & \times \left(\frac{\mathit{a}}{0.1\, \SI{}{\micro\meter}}\right)^{-1} \left(\frac{\mathit{n}_{\rm H}}{100\, \SI{}{\centi\meter^{-3}}}\right)^{-1} \\
                 & \times \left(\frac{\mathit{T}_{\rm gas}}{20\, \SI{}{\kelvin}}\right)^{-1/2} \left(\frac{\mathit{Z}_{\rm gas}}{0.02}\right)^{-1} [yr],
\end{aligned}
\end{equation}
where the first term defines the reference timing of growth activation, $a$ is a dust grain size with a typical size of $\sim \SI{0.1}{\micro\meter}$ (\citealt{liqi19}). The $n_{\rm H}$ and $T_{\rm gas}$ are the number density and temperature of the ISM gas respectively, and $Z_{\rm gas}$ is a gas-phase metallicity\footnote{Note that given the \texttt{SIMBA}’s $\sim\rm kpc$ resolution, a multiphase galaxy ISM is not resolved, which prevent us from knowing the exact dependence of the modelled dust content to the gas surface density}. Molecular gas number density is calculated from the surface gas density following \cite{popping17}. We calculate the gas surface density using the gas mass and the half-mass gas radius from \texttt{SIMBA} together with the metallicity. We find that dust re-growth time is the most rapid amongst the slowly quenched dust-rich QGs of group B ($\tau_{\rm acc}=70.64^{+66.98}_{-42.2}$ Gyr). For the fast QGs of group A and the slow dust-poor QGs of group C, typical re-growth timescales are slightly longer on average.

\begin{table*}[ht]
    \centering
    \renewcommand{\arraystretch}{2}
    \scalebox{0.84}{\begin{tabular}{lccc}
        \hline
        \multirow{2}{3.5cm}{\textbf{ }} & \textbf{Fast} & \multicolumn{2}{c}{\textbf{Slow}}\\
        \hline
        & \textbf{ } & $\bm{\log(f_{\rm dust})>-4}$ & $\bm{\log(f_{\rm dust})<-4$}\\
        \multicolumn{4}{c}{\textbf{Stellar properties}} \\
        \hline
        \textbf{$\Delta_{\rm MS}$} & \(-1.4^{+0.24}_{-0.18}\) & \(-1.37^{+0.27}_{-0.25}\) & \(-1.44^{+0.35}_{-0.2}\) \\
        \textbf{$\log(M_{{\star}}/M_{\odot})$} & \(10.23^{+0.4}_{-0.27}\) & \(9.8^{+1.08}_{-0.39}\) & \(10.45^{+0.31}_{-0.24}\) \\
        \textbf{$ \rm Age [Gyr]$} & \(1.69^{+1.55}_{-0.56}\) & \(3.92^{+1.56}_{-1.7}\) & \(2.24^{+1.63}_{-0.76}\) \\

        \multicolumn{4}{c}{\textbf{ISM properties}} \\
        \hline
        \textbf{$\log(f_{{\rm dust}})$} & \(-3.09^{+0.27}_{-0.55}\) & \(-3.29^{+0.27}_{-0.39}\) & \(-5.13^{+0.7}_{-0.62}\) \\
        \textbf{$\log(f_{{\rm H_2}})$} & \(-1.37^{+0.2}_{-0.3}\) & \(-1.37^{+0.17}_{-0.17}\) & \(-1.2^{+0.13}_{-0.14}\) \\
        \textbf{$\log(\delta_{{\rm DGR}})$} & \(-1.72^{+0.37}_{-0.52}\) & \(-1.89^{+0.25}_{-0.45}\) & \(-3.95^{+0.81}_{-0.64}\) \\
        \textbf{$12+\log(\rm O/H)$} & \(8.79^{+0.12}_{-0.12}\) & \(8.73^{+0.26}_{-0.18}\) & \(8.96^{+0.07}_{-0.08}\) \\
        \textbf{$\log(\delta_{{\rm DTM}})$} & \(-0.27^{+0.15}_{-0.39}\) & \(-0.31^{+0.09}_{-0.42}\) & \(-2.35^{+0.79}_{-0.64}\) \\
        
        \multicolumn{4}{c}{\textbf{Timescales}} \\
        \hline
        \textbf{$t_{{\rm quench}} [\rm Gyr]$} & \(0.05^{+0.05}_{-0.02}\) & \(0.82^{+0.98}_{-0.48}\) & \(0.61^{+0.7}_{-0.28}\) \\
        \textbf{$t^{{H_2}}_{{\rm removal}} [\rm Gyr]$} & \(5.73^{+2.18}_{-3.0}\) & \(2.81^{+4.22}_{-1.66}\) & \(6.42^{+1.8}_{-2.54}\) \\
        \textbf{$t_{{\rm acc}} [Myr]$} & \(87.75^{+28.7}_{-28.94}\) & \(70.64^{+66.98}_{-42.2}\) & \(86.66^{+30.82}_{-27.97}\) \\

        \bottomrule
        \bottomrule
    \end{tabular}
    }

    \caption{Medians and associated uncertainties (16th–84th percentile range) of various physical properties related to fast and slow quenching QGs. The properties are computed for QGs corresponding to the three regions in Fig.~\ref{fig:f_dust_age_exponential_evol} at the quenching phase (see also Fig.~\ref{fig:scketch}). Note that for the slow quenchers we made additional division based on $f_{\rm dust}$ to reflect the difference between the two clouds as illustrated in Fig.~\ref{fig:f_dust_age_exponential_evol}.}
    \label{tab:examples_values}
\end{table*}

The re-growth timescales obtained for all three groups of QGs are still sufficiently quick to compete against the sputtering timescales estimated for our dusty QGs ($\gtrsim10^{8}\rm yr$). It is feasible for dust grains to re-form after the swept-up ISM has cooled. As pointed out by \cite{hirashita2017dust}, as long as sputtering timescales are longer than the AGN cycle, and with short-enough re-growth timescales, there would be enough residual material to reform the dust on ISM metals. \cite{donevski23} propose the grain growth mechanism as an plausible explanation for the observed different pathways of early-type QGs, for which more than 500 Myr elapsed after the quenching. Timescales for such process counteracting the dust destruction are expected to be short, of the order of $50-250\:\rm Myr$ (\citealp{hirashita2015dust}). Since the timescales for $\rm H_{2}$ gas removal in group B are similar to that for dust destruction, the presence of large $f_{\rm dust}$ in such galaxies at the dry-phase strongly supports the idea of dust re-growth happening in $< \SI{500}{\mega\year}$. \cite{Ni2022} investigated the relation between the observed fraction of X-dominated AGNs and their stellar age in the COSMOS field. They found a strong anti-correlation, with the oldest objects having lower AGN accretion rates. They concluded that a significant fraction of the accretion on the AGN is due to hot recycled gas, which cools and affects the AGN on a timescale of $\sim \SI{1}{Gyr}$. This means that, after the reduction of $H_2$ gas from the quenching, the ISM would have enough time before the re-activation of an intense X-ray AGN feedback to undergo dust re-growth, in qualitative agreement with \texttt{SIMBA} predictions.

It is important to note that prolonged dust growth is not the sole process influencing ISM dust abundance and it does not rule out contribution of AGB stars. Dust growth through accretion on ISM metals requires seed grains to be deposited into the ISM for which AGB stars are likely source (see e.g., \citealt{nanni13}). An alternative possibility, as proposed in several observational works, suggests a dust mixing via interaction of AGN winds with star formation-driven outflows. The latter effect seem to be particularly relevant for lower-mass galaxies (see e.g., \citealt{romano23}). Investigating this possibility is out of scope of this study.

We recover the same dust-poor but gas-rich QGs (group C) characterised with exotically low $\delta_{\rm DGR}$ that are first identified by \cite{whitaker21a}. The presence of such QGs can be interpret by mean of dilution and inefficient dust grain growth, as already proposed by \cite{hirashita2017dust}). In this scenario, during a long-lasting ($\sim5-6$ Gyr) timescales, radiative-mode AGN feedback steadily removes the $\rm H_{2}$ gas, and dust grains become efficiently destroyed by thermal sputtering and do not have a way to withstand the erosion by growing through condensation of metals. On top of this, inflows of dust-poor cold gas would decrease the $\delta_{\rm DGR}$, keeping the group C sources in the lower part of Fig.~\ref{fig:DGR_plane_evol}. It is worth noting that for the case of inefficient dust growth, their $\delta_{\rm DGR}$ becomes highly sensitive to the $M_{\star}$, and strongly affects most massive QGs, as we show in Section~\ref{ISM dust-gas abundance within the MS}.

To further explore dust growth scenario we analyse the dust-to-metal ratio ($\delta_{\rm DTM}$) which is proven to be a great proxy for the dust re-growth (see e.g., \citealt{de2019systematic}). We follow \cite{de2019systematic} and calculate $\delta_{\rm DTM}$ as $\delta_{\rm DTM}$ = $M_{\rm Z}$/$M_{\rm d}$ where $M_{\rm Z}$ = $f_{\rm z} \times {\rm M}_{\rm g} + {\rm M}_{\rm d}$ is the total mass of metals, $M_{\rm d}$ is the dust mass, $M_{\rm g}$ is the gas mass and $f_{\rm Z}=27.36 \times 10^{12+{\rm log}_{10}({\rm O/H})-12}$ is the mass fraction of metals. Values are listed in Tab.~\ref{tab:examples_values}. We find that at the end of the quenching phase, both group A and B have large values of $\log(\delta_{\rm DTM}) > -0.8$, while for group C the value is at least twice as big. The values for group A and B are comparable with the literature reference $\log(\delta_{\rm DTM}) \sim -0.52$ (i.e. \citealp{2015Camps}, \citealp{2016Clark}) and the values from DustPedia galaxies ($\log(\delta_{\rm DTM}) \sim -0.67$; \citealt{de2019systematic}) measured for QGs. At the gas-removed evolutionary point the scenario is rather different, with median of $\log(\delta_{\rm DTM, A}) = -2.82^{+2.46}_{-0.81}$, $\log(\delta_{\rm DTM, B}) = -0.27^{+0.1}_{-1.16}$, $\log(\delta_{\rm DTM, C}) = -3.14^{+0.89}_{-0.69}$. While there is a significant decrease in $\delta_{\rm DTM}$ for both groups A and C, the value for group B stays rather high up to the dry-phase, indicating the presence of dust growth in a large part of the sample. We qualitatively summarize our results in Fig.~\ref{fig:scketch}, where the properties of the groups A, B, and C are shown in relation to $f_{\rm dust}$ and stellar age, together with the direction of evolution and dust growth.

\subsection{Observational implications for studying high-$z$ dusty QGs}

While it has been commonly assumed that QGs are dust-poor, the presence of significant dust content in \texttt{SIMBA} QGs predicted by this study imposes several important implications. Firstly, the substantial ISM dust abundance can severally affect the intrinsic SEDs of QGs and selection techniques based on rest-frame colours (see e.g., \citealt{akins2022quenching}, \citealt{long23}, \citealt{setton24}). Secondly, we show that observed dust and cold gas tightly follow each other until the end of a quenching phase. After galaxy becomes fully quenched, the observed $f_{\rm dust}$ may be resulting from a dust growth, with re-formation timescales which may be offsetting from those attributed to $\rm H_{2}$ gas exhaustion. For that reason, we urge caution when using $f_{\rm dust}$ to convert to $f_{\rm gas}$ and ultimately to $M_{\rm gas}$ in QGs. This seems to be particularly important for planning the ALMA observations to probe the cold gas content via commonly used tracers such as CO transitions. At a given MS-offset (or stellar age), older and metal-richer QGs with enhanced $f_{\rm dust}$ would require longer integration times for the detection of molecular gas than those simply extrapolated from the typical $\delta_{\mathrm{DGR}}\sim1/100$. This is a consequence of enhanced $\delta_{\mathrm{DGR}}$ and subsequently lower $M_{\rm gas}$. Indeed, recent studies pointed out to alternative tracers such as [CII] 158 $\mu$m emission line to be more efficient than CO for probing a gaseous component of QGs at $z>2$ (\citealt{deugenio23}). We thoroughly investigate the two above-mentioned challenges in our accompanying study (Lorenzon et al., in prep.). Finally, our findings suggests the existence of dust-rich QGs at $z\sim1-2$ that were quenched $>5$ Gyr ago ($z_{\rm q}\gtrsim3-5$) through the rapid and intense AGN activity, similar to those that were confirmed observationally with JWST (\citealt{stacey22}, \citealt{kocevski23}, \citealt{deugenio24}) and with ALMA (\citealt{salak24}). Joint observational strategies between JWST and ALMA are required to test this specific prediction. If confirmed, such rapidly quenched but dust-rich galaxies will provide a unique insight into the SMBH-ISM co-evolution in the evolved objects caught in the first 1-2 Gyr of the Universe's history.

\section{Conclusions} \label{conclusion}

We use the full suite of \texttt{SIMBA} cosmological simulation and provide the theoretical framework for the pathways of dust and cold gas evolution in QGs at high-$z$. We identify a mass-complete sample ($M_{\star}>10^9M_{\odot}$) of QGs that are quenched over a wide redshift range ($0 < z \lesssim2$) and study their ISM dust and gas abundance at different critical points in evolutionary history. We account for their large-scale environments, exploring separately the QGs residing in the field and in clusters.

Our main results are summarised as follows: 
\begin{itemize}
    
    \item Quenching timescales in \texttt{\texttt{SIMBA}} are strongly bimodal, and no prevalence of a specific quenching mode has been found for a given environment. The bimodality does not translate directly into dust-rich and dust-poor galaxies, suggesting that the quenching mode alone does not have a prevalent effect on the ISM evolution. 
	
    \item The $\delta_{\rm DGR}$-$Z_{\rm gas}$ plane for QGs consists of two distinct regions. Roughly half of QGs follow the relation expected for SFGs and the other half exhibits the absence of a clear trend accompanied by a large scatter ($\sim4$ orders of magnitude) at super-solar $Z_{\rm gas}$. Such behavior suggests the diversity of efficiencies for the production and removal of dust and gas in QGs. Although the overall shape remains the same for the field and cluster QGs, we find that dust destruction is enhanced by sputtering in cluster environments.

    \item The stellar masses in the quenching phase have a profound impact on further dust/gas evolution in QGs. As they depart from the MS towards the final gas-dry phase ($f_{\rm gas}<0.1\%$), more than a half of QGs with intermediate masses ($\log(M_{\star}/M_{\odot})\lesssim10.5$) sustain high ISM dust abundance ($f_{\rm dust}>10^{-3}-10^{-4}$; $\delta_{\rm DGR}>1/100$). This is in contrast to massive QGs ($10.5<\log(M_{\star}/M_{\odot})<11.25$) that experience a big drop (>1-2 orders of magnitude) in their $f_{\rm dust}$ and $\delta_{\rm DGR}$. 
    
    \item Enhanced $f_{\rm dust}$ and $\delta_{\rm DGR}$ are mostly a consequence of internal processes, and only marginally of external effects such as mergers and rejuvenations. The dominant dust re-formation mechanism is prolonged dust growth in metal-rich ISM. This key process is effective in the first $100-500$ Myr of a quenching phase, allowing QGs to extend the dust destruction timescales expected for thermal gas heating.

    \item We successfully reproduce the observational relation of $f_{\rm dust}$ with stellar age in QGs identified at $0<z\lesssim2$. We interpret the emergence of these sources as a mix of both fast and slow quenchers, with fast quenchers predominantly having younger stellar ages (<3 Gyr) and slow quenchers dominating the large scatter towards older stellar ages (>3–10 Gyr). During the quenching phases, the bulk of QGs show little to no evolution in their $f_{\rm dust}$ with age. This implies long dust depletion timescales (>2-3 Gyr), a direct consequence of prolonged grain growth. 
    
    \item Overall, we predict that for significant number of QGs, removal timescales for the cold $\rm H_{2}$ gas and dust are mutually different, owing to a complex interplay of quenching pathways attributed to the AGN, and various dust re-growth efficiencies. This prediction may have important implications for characterising the SMBH-ISM co-evolution in QGs. 
\end{itemize}

We stress that the aforementioned predictions urge immediate observational investigations of dusty QGs. In particular, to better understand the strength of the AGN activity and its coupling to the ISM dust and gas, our study suggests a systematic search and analysis of ISM in high-$z$ QGs of low and intermediate masses ($9<\log(M_{\star}/M_{\odot})<10.5$). Although challenging, this observational quest is achievable with the current generation of instruments such as JWST, ALMA and \textit{Euclid}. The JWST MIRI, in particular, can be used to probe the warm dust and molecular gas through MIR diagnostics in individual dusty QGs at various cosmic epochs up to $z\sim 2-3$, while its combination with NIRspec will help access the level of AGN activity and better constrain gas-phase metallicity. Synergy with sub-mm instruments (ALMA, NOEMA) and near-IR survey telescopes (\textit{Euclid}) is needed to infer the census on the cold dust and large-scale environments of selected QGs, respectively. In our complementary work, we provide the recipes for such observational tests and present the key differences in SEDs of dust-rich versus dust-poor QGs that were formed at different cosmic epochs (Lorenzon et al., in prep).

\begin{acknowledgements}
G.L.and D.D acknowledge support from the National Science Center (NCN) grant SONATA (UMO-2020/39/D/ST9/00720). K.L acknowledges support of the Polish Ministry of Education and Science through the grant PN/01/0034/2022 under `Perły Nauki' program.
AWSM acknowledges the support of the Natural Sciences and Engineering Research Council of Canada (NSERC) through grant reference number RGPIN-2021-03046. J is grateful for support from the Polish National Science Centre via grant UMO-2018/30/E/ST9/00082.
A.N, and M.R. acknowledge support from the Narodowe Centrum Nauki (NCN), Poland, through the SONATA BIS grant UMO2020/38/E/ST9/00077
\end{acknowledgements}

\bibliographystyle{aa} 
\bibliography{bibliography} 

\onecolumn

\begin{appendix} 

\section{Evolution of the $\delta_{\rm DGR}$-metallicity plane with stellar mass} \label{appendix1}

\begin{figure*}[tbh!]
    \centering
    \includegraphics[width=16.4cm,clip]{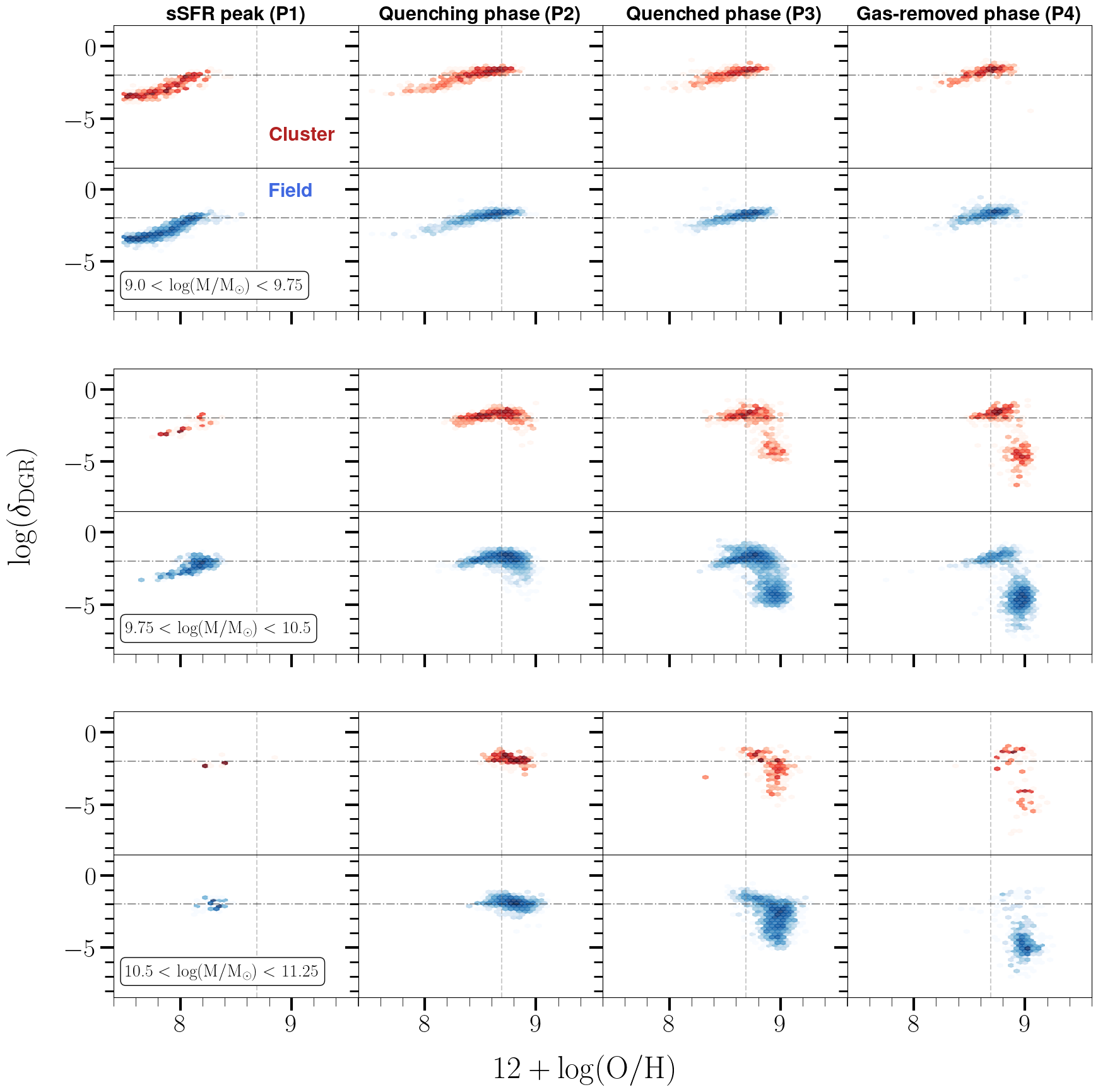}
    \caption{The evolution of the $\delta_{\rm DGR}$-metallicity plane according to the four evolutionary points (from left to right) introduced in Sect.~\ref{Evolution stages} and the same mass bins used in Fig.~\ref{fig:DGR_plane_evol} (from top to bottom). Galaxies from the cluster sample are represented in red, while galaxies from the field sample are in blue. Vertical gray dotted lines show the solar value for the metallicity ($12+\log(O/H) \equiv 8.69$). Horizontal gray dashed lines show the literature reference value of $\delta_{\rm DGR} \equiv 1/100$.}
    \label{appfig:mass_distrib}
\end{figure*}

Fig.~\ref{appfig:mass_distrib} shows the $\delta_{\rm DGR}$-metallicity plane at the evolutionary points defined in Sect.~\ref{Evolution stages}, evolving from SF to completely passive (left to right) and from low to high $M_{\star}$ (top to bottom) for both cluster (red) and field (blue) samples. Galaxies are selected with $\log(M_{\star}/M_{\odot}) > 9$ in order to achieve completeness of the sample (\citealt{tacconi2018phibss}, \citealt{ghodsi2023star}). It is clear that the behavior of both cluster and field samples is qualitative similar. As the low mass ($\log(M_{\star}/M_{\odot}) < 9.75$) galaxies evolve, they maintain a quite large $\delta_{\rm DGR}$. The intermediate mass range ($ 9.75 < \log(M_{\star}/M_{\odot}) < 10.5$) presents a large scatter of $\delta_{\rm DGR}$ that appears during the quenching of the galaxies (between P2 and P3). Even for super-solar metallicities, galaxies can maintain $\delta_{\rm DGR}$ comparable to the low mass interval. Galaxies with the highest metallicities have systematically lower $\delta_{\rm DGR}$. For the range of large masses ($10.5 < \log(M_{\star}/M_{\odot}) < 11.25$) metallicities are systematically higher at all evolutionary points. The scatter in P3 is less prominent with respect to the sample of intermediate masses, but there are less galaxies with large $\delta_{\rm DGR}$. A very few galaxies have measurable metallicity, $M_{H_2}$ and $M_{\rm dust}$ in P4, and they mostly populate the low $\delta_{\rm DGR}$ region of the plane. At this stage, field galaxies more dominantly populate the $\log(\delta_{\rm DGR}) < -2$ region with respect to cluster galaxies, that are sparsely found with both $\log(\delta_{\rm DGR}) < -2$ and $\log(\delta_{\rm DGR}) > -2$.

\section{Evolutionary tracks for representative QGs} \label{appendix2}

We show in Fig.~\ref{appendix2} the evolution of the four representative field galaxies (right panel) from Fig.~\ref{fig:SFH_examples} on the dust-stellar-age plane. Three out of the four galaxies (G1, G3, G4) are in the region of high $f_{dust}$ despite having different evolutionary paths. Such galaxies start the quenching process with an older stellar population than G2. G1 increases its dust content up to the quenching phase and then it decreases it. The $t_{\rm quench}$ is the longest in the sample. The $f_{\rm dust}$ at $z=0$ is below the resolution for \texttt{SIMBA}, thus is not present in the figure. G2 is the only galaxy ending in the bulk of low dust fraction, after a long and continuous evolution. G2 has the longest $t^{H_2}_{\rm removal}$, one order of magnitude larger than G1, and the second longest $t_{\rm quench}$ of the four galaxies. The $f_{\rm dust}$ of G2 at $z=0$ is low, but still above the resolution limit for \texttt{SIMBA}. Moreover, G2 is the only galaxy of the sample experiencing the jet-dominated AGN feedback mode during the quenching phase. G3 is the only fast quencher of the four galaxies. G4 has the shortest $t^{H_2}_{\rm removal}$, one order of magnitude lower than G1 and G3 and two orders of magnitude lower than G2, and its $f_{\rm dust}$ increases at $z=0$. It is important to notice that, despite the longer $t^{H_2}_{\rm removal}$, G2 end up with a larger $f_{\rm dust}$ than G1 and G3 at $z=0$, suggesting that the timescale by itself is not enough to predict the evolution of the dust content. 

The situation is similar for the cluster galaxies (left panel), where the galaxy experiencing jet-mode feedback (G6) reaches lower values of $f_{\rm dust}$ at the dry-phase. G6 has a very short $t^{H_2}_{\rm removal}$, and a relatively low $t_{quench}$. The majority of the dust content is lost during the quenching phase, cold gas is lost soon after and the dust content rapidly follows with no measurable dust at $z=0$. The galaxy G5 follows the same trend observed for G1, having also a similar SFH.

\begin{figure*}[tbh!]
    \centering
    \includegraphics[width=16.4cm,clip]{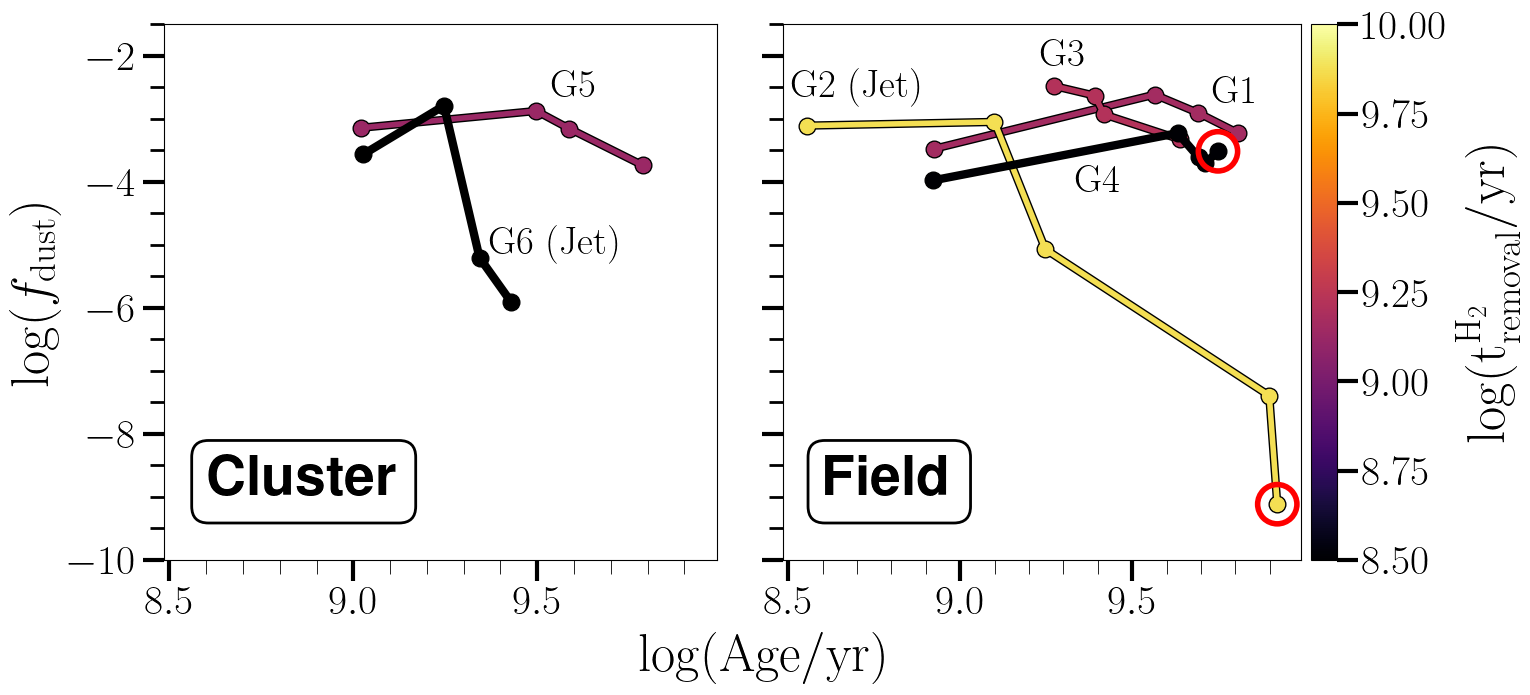}
    \caption{The evolution of galaxies from Fig.~\ref{fig:SFH_examples} in the dust-stellar-age plane for the cluster (left) and field (right) galaxies. The mass weighted stellar age is on the x-axis, while the $f_{\rm dust}$ is shown on the y-axis. Each galaxy is color coded with the timescale $t^{H_2}_{\rm removal}$ to reach the dry-phase (P4) from the quenched phase (P3). The properties of each galaxy are evaluated from left to right at the P1, P2, P3 and P4. If present, the red circle symbolize the values at $z=0$.}
    \label{appfig:dust_age_example_both}
\end{figure*}

\end{appendix}
\end{document}